\documentclass[prl,showpacs,floatfix,nontitlepage,twocolumn,superscriptaddress]{revtex4-1}
\usepackage[colorlinks=true, citecolor=blue, urlcolor = blue, linkcolor= red, 
bookmarks=true]{hyperref}
\usepackage{natbib}
\usepackage{amsmath}    
\usepackage{graphicx}   
\usepackage{verbatim}   
\usepackage{color}      
\usepackage{subfloat}
\usepackage{hyperref}   
\usepackage{graphicx, psfrag}
\usepackage[centerlast]{caption}
\usepackage{float}
\usepackage[subrefformat=parens,labelformat=parens]{subfig}
\usepackage{siunitx}

\newcommand{\ceri}{$\text{Ce}_2\text{O}_3$}

\begin{document}

\title{Emergent Vibronic Excitations in the Magnetodielectric Regime of \ceri: Raman Scattering Studies}

\author{A. Sethi}
\email{asethi8@illinois.edu}
\affiliation{Department of Physics and Frederick Seitz Materials Research Laboratory, University of Illinois, Urbana, Illinois 61801, USA}

\author{J. E. Slimak} 
\affiliation{Department of Physics and Frederick Seitz Materials Research Laboratory, University of Illinois, Urbana, Illinois 61801, USA}

\author{T. Kolodiazhnyi}
\affiliation{National Institute for Materials Science, 1-1 Namiki, Tsukuba, Ibaraki, 305-0044, Japan}

\author{S. L. Cooper}
\email{slcooper@illinois.edu}
\affiliation{Department of Physics and Frederick Seitz Materials Research Laboratory, University of Illinois, Urbana, Illinois 61801, USA}

\begin{abstract}

 The strong coupling between spin, lattice and electronic degrees of freedom in magnetic materials can produce interesting phenomena, including multiferroic and magnetodielectric (MD) behavior, and exotic coupled excitations, such as electromagnons. We present a temperature- and magnetic-field-dependent inelastic light (Raman) scattering study that reveals the emergence of vibronic modes, i.e., coupled vibrational and crystal-electric-field (CEF) electronic excitations, in the unconventional rare-earth MD material, \ceri. The energies and intensities of these emergent vibronic modes are indicative of enhanced vibronic coupling and increased modulation of the dielectric susceptibility in the N\'eel state ($T_\text{N} \approx 6.2\,\text{K}$). The field-dependences of the energies and intensities of these vibronic modes are consistent with a decrease of both the vibronic coupling and the dielectric fluctuations associated with these modes below $T_\text{N}$. These results suggest a distinctive mechanism for MD behavior in \ceri\,that is associated with a field-tunable coupling between CEF and phonon states.
 
\end{abstract}

\maketitle

\textit{Introduction}-- Magnetically responsive materials, including multiferroics~\cite{Che-2007} and magnetodielectrics~\cite{Muf-2010}, are of enormous scientific and technological interest because the physical properties of these materials have large susceptibilities to external perturbations, such as magnetic field and pressure. These large susceptibilities reflect a strong coupling between spin, electronic, and lattice degrees of freedom, one manifestation of which is the emergence of coupled excitations, such as hybrid magnon-phonon modes (i.e., “electromagnons”) in the magnetoelectric phase of multiferroic materials~\cite{Che-2007, Agu-2009, Sus-2007, Pim-2006, Don-2015, Ort-2015, Jur-2017, Kat-2007}. A key challenge remains elucidating the underlying spin-lattice coupling mechanisms governing magnetically responsive phases and their associated hybrid excitations~\cite{Che-2007}.

Multiferroic and magnetodielectric materials most commonly involve transition metal (TM) compounds, in which the presence of magnetic frustration and the resultant complex magnetic orders are key factors responsible for their multiferroic properties~\cite{Che-2007, Muf-2010,Agu-2009, Sus-2007, Jur-2017,Pim-2006, Kat-2007}. The search for new magnetically responsive materials is important, not only to expand the family of existing multiferroics but also to uncover the novel physics that new materials may exhibit. 

Recently, a giant MD response was reported in the hexagonal A-type rare-earth oxide, \ceri\,\cite{Kol-2018}. This material does not exhibit magnetic frustration or non-collinear magnetic order typical of magnetically responsive TM compounds; consequently, the mechanism associated with MD behavior in \ceri\,remains an open issue. Significantly, the localized f-orbitals in rare-earth materials lead to CEF excitations within the f-electron manifold that are energetically resonant with the phonon bands~\cite{Gru-2002,Tha-1982, Loe-2003, Gau-2018, Adj-2012, Hey-1991}. This condition can lead to strong vibronic coupling between the CEF and phonon excitations~\cite{Tha-1982, Loe-2003, Gau-2018, Adj-2012, Hey-1991}, offering the possibility of novel spin-lattice coupling mechanisms and magnetically responsive phenomena in rare-earth-based materials.

In this paper, we report Raman scattering studies showing that the MD phase below $T_\text{N}$ in \ceri\,is characterized by the emergence of hybrid ''vibronic'' excitations, involving coupled f-electron CEF and phonon modes. The energies and intensities of these vibronic modes are consistent with a rapid enhancement of the vibronic coupling and an increased modulation of the dielectric susceptibility by these coupled modes below $T_\text{N}$ in \ceri. Magnetic-field-dependent Raman measurements show that an applied field decreases both the vibronic coupling and the modulation of the dielectric susceptibility below $T_\text{N}$. These results suggest that the field-dependent coupling between the electronic and phononic degrees of freedom in the N\'eel state is associated with a distinctive mechanism for MD behavior in \ceri.

\textit{Experimental Methods}-- The polycrystalline \ceri\,samples used for this study were prepared and characterized as per the methods described in Ref.~\cite{Kol-2018}. Raman scattering measurements were performed using a $647.1\,\text{nm}$ excitation line of a Kr$^+$ laser. To minimize laser heating of the samples, the laser was focused to a $\sim \text{\SI{50}{\micro\meter}}$ diameter spot and incident powers between $\text{\SI{0.5}{\milli\watt}}$ and $\text{\SI{3}{\milli\watt}}$ were used.  
The minimal effects of laser heating at these low laser powers were estimated using a procedure discussed in the Supplemental Material, and these quantitative estimates are accounted for in the temperatures reported in this paper. The scattered light was collected in a back-scattering geometry, dispersed through a triple stage spectrometer and then recorded with a liquid nitrogen cooled CCD detector. Samples were inserted in a helium flow-through cryostat, which was horizontally mounted in the bore of a superconducting magnet to allow simultaneous temperature ($3 - 300$ K) and magnetic ($0 - 9$ T) measurements. All magnetic measurements were performed in a Faraday geometry with the incident light wave vector parallel to the magnetic field ($\boldsymbol k  \parallel \boldsymbol H$) and perpendicular to the surface of the sample. The incident light was circularly polarized and the scattered light from all polarizations was collected.

\begin{figure}
\subfloat{\includegraphics{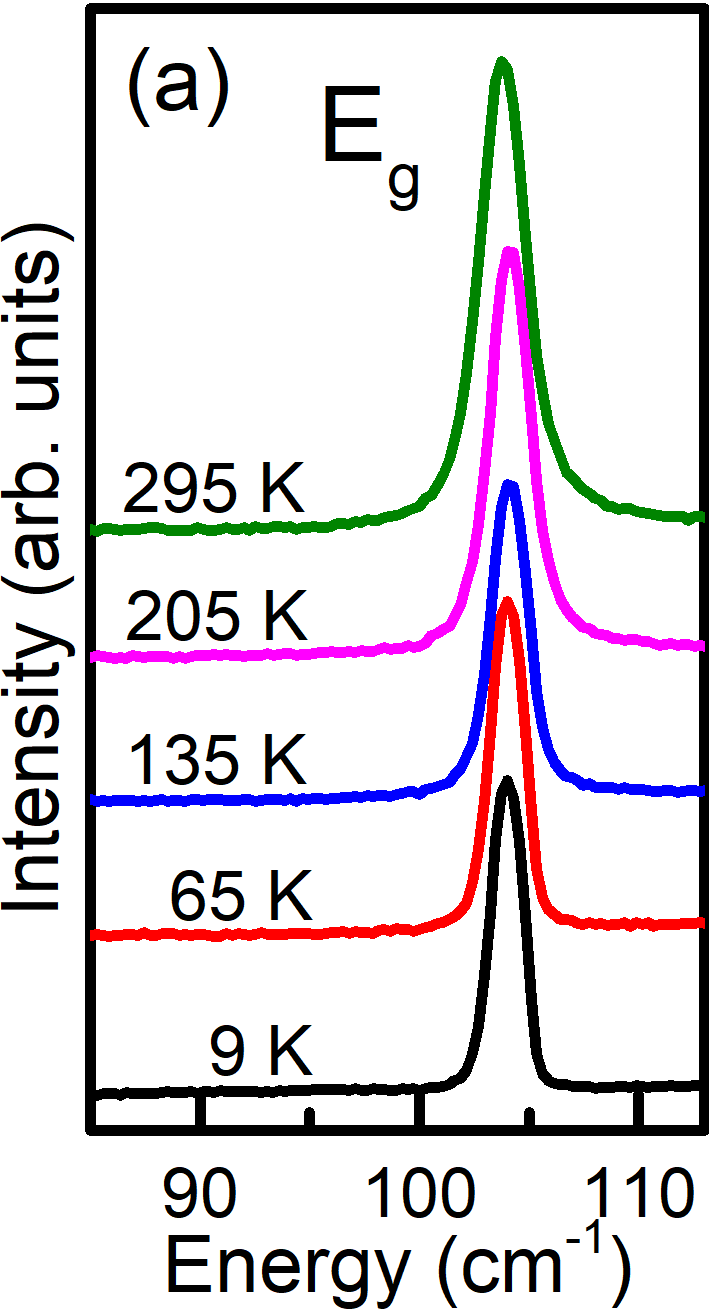} \label{Fig1a}}
\subfloat{\includegraphics{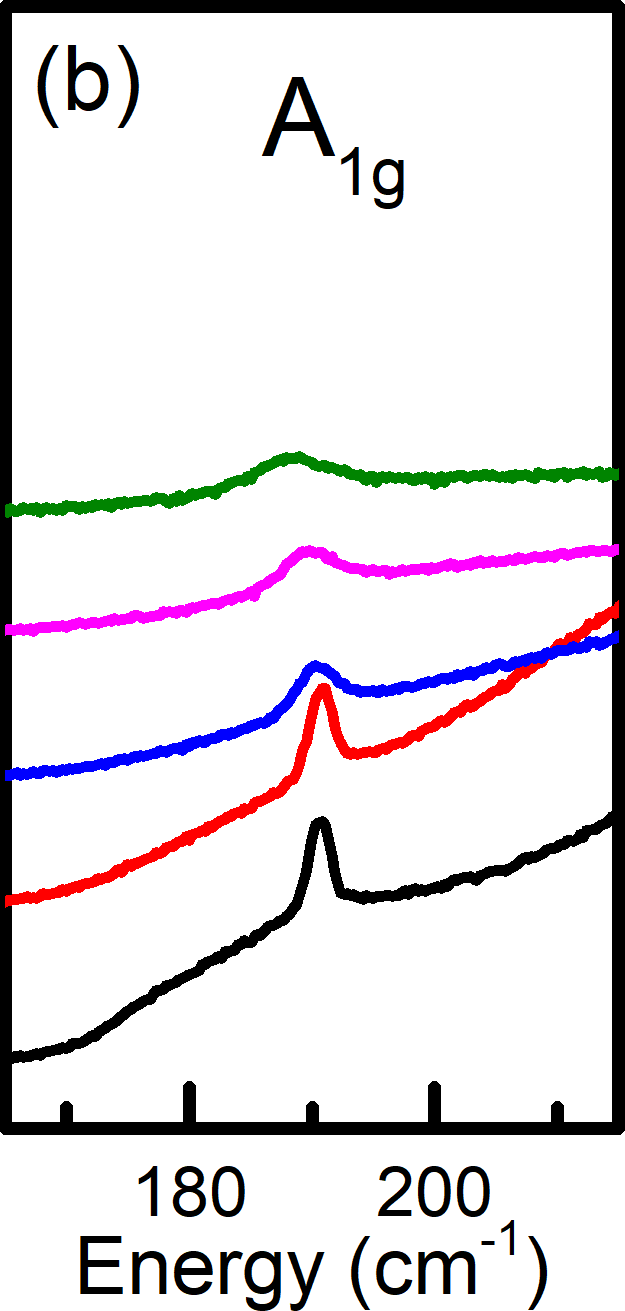} \label{Fig1b}} 
\subfloat{\includegraphics{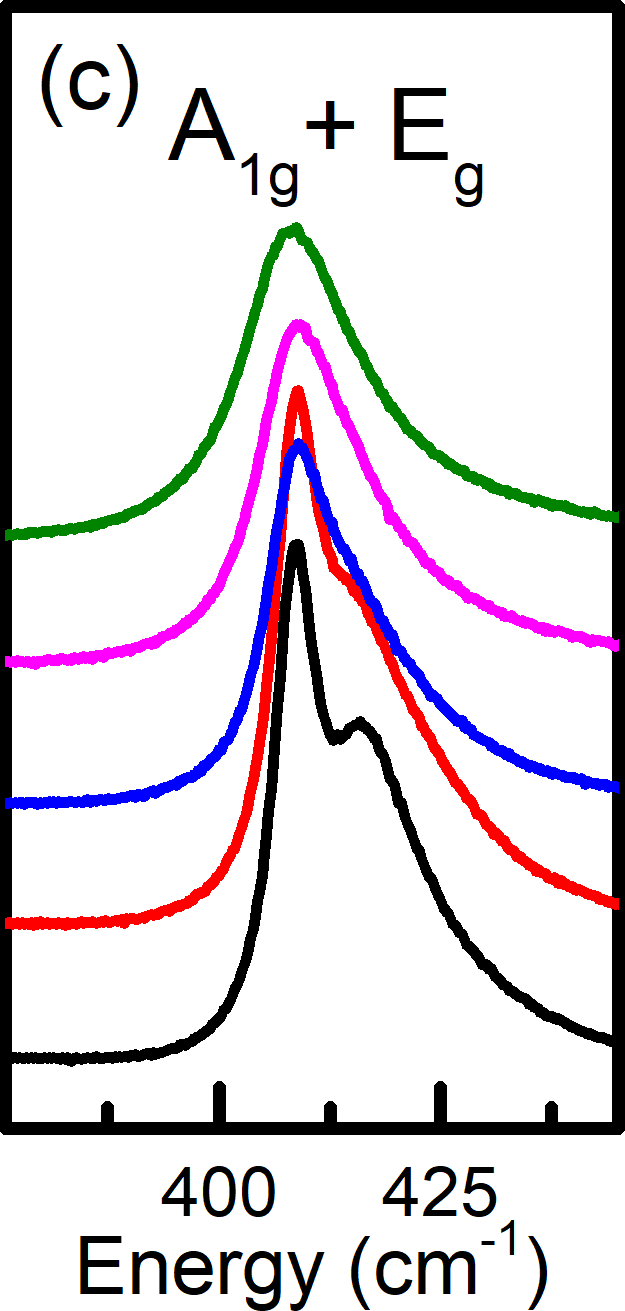} \label{Fig1c}}

\caption{\label{Fig. 1}Temperature dependences of Raman intensities of zone-center phonons for $T > T_\text{N}$.}
\end{figure}

\textit{Phonons for $T > T_\text{N}$} -- \textit{A}-type \ceri\,belongs to the trigonal P$\bar{3}2$/m1 or D$_{3d}^3$ space group~\cite{Kol-2018, Avi-2016, Gou-1981, Zar-1979}. Group theory predicts four Raman active phonons for this crystal structure, $2A_{1g} + 2E_g$~\cite{Avi-2016, Gou-1981, Zar-1979}. Figure~\ref{Fig. 1} shows the temperature dependences of the expected phonons in the paramagnetic phase of \ceri. The measured phonon energies at $T=295 \,\text{K}$ are in close agreement with a previous room temperature Raman scattering study~\cite{Avi-2016}. The two lower energy phonons shown in Figs.~\ref{Fig1a}--\ref{Fig1b} at $\omega \approx 103.8\,\text{and}\,188.5\,\text{cm}^{-1}$ correspond to the bending vibrations of the Ce-O(II) bond with $E_g$ and $A_{1g}$ symmetry, respectively~\cite{Avi-2016, Gou-1981}. The increasing background of the phonon in Fig.~\ref{Fig1b} is caused by a nearby CEF excitation that will be discussed later. The broad peak in Fig.~\ref{Fig1c} at $\omega \approx 408\,\text{cm}^{-1}$ (at $T=295$~K) is comprised of two closely spaced $A_{1g}$ - and $E_g$ - symmetry phonons, which are associated with the stretching vibrations of the long and short Ce-O(II) bonds, respectively~\cite{Avi-2016, Gou-1981, Zar-1979}. In unpolarized Raman scattering studies at room temperature, the stretching $A_{1g}$ and $E_g$ phonons were not energy resolved in previous studies of the rare-earth (RE) sesquioxides (RE$_2$O$_3$), including \ceri~\cite{Avi-2016, Gou-1981}. However, at $9$ K, we can clearly distinguish both the phonons at $\omega_{A_{1g}}\approx408.6\,\text{and}\,\omega_{E_g}\approx416\,\text{cm}^{-1}$.

\begin{figure}
\subfloat{\includegraphics{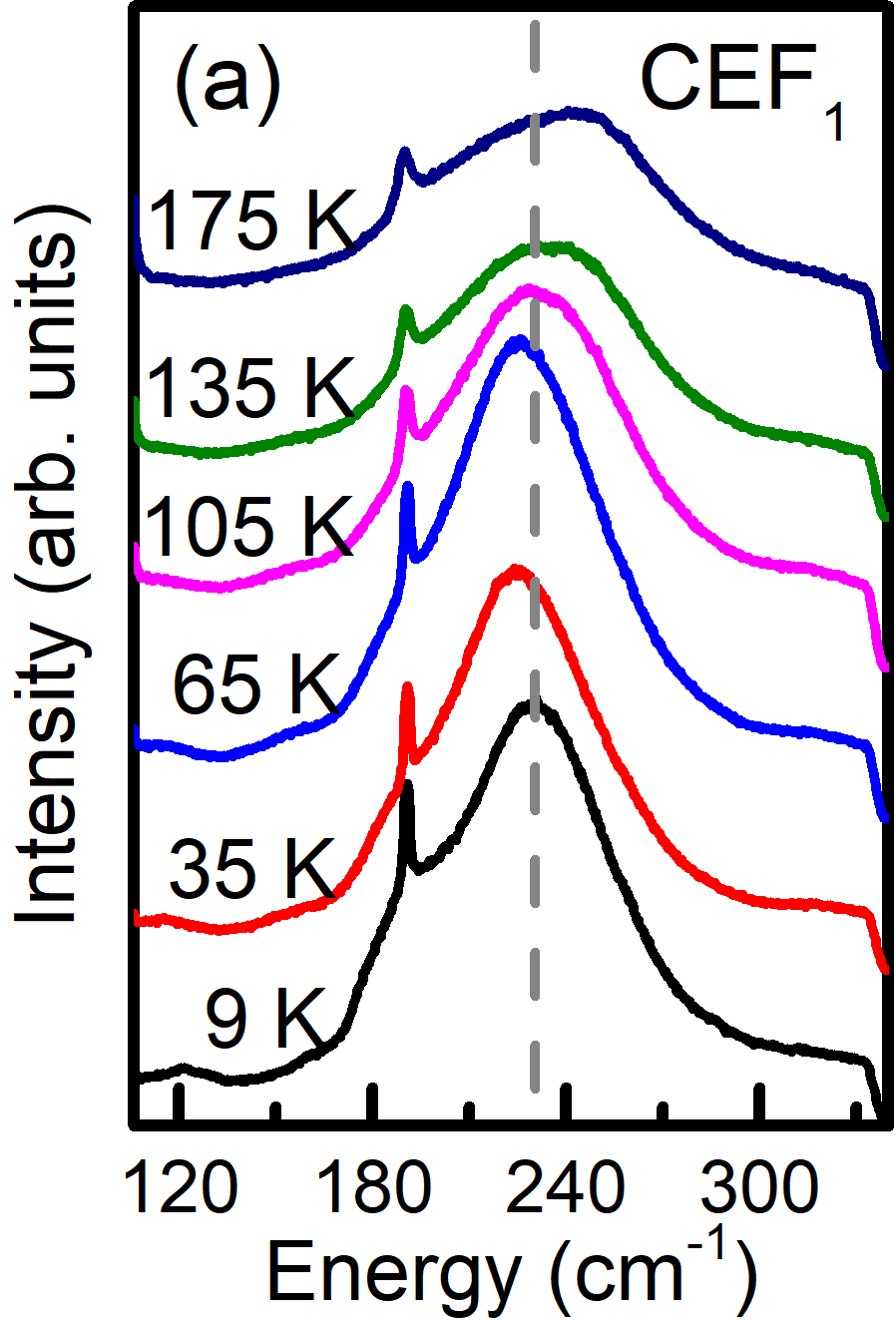} \label{Fig2a}}
\subfloat{\includegraphics{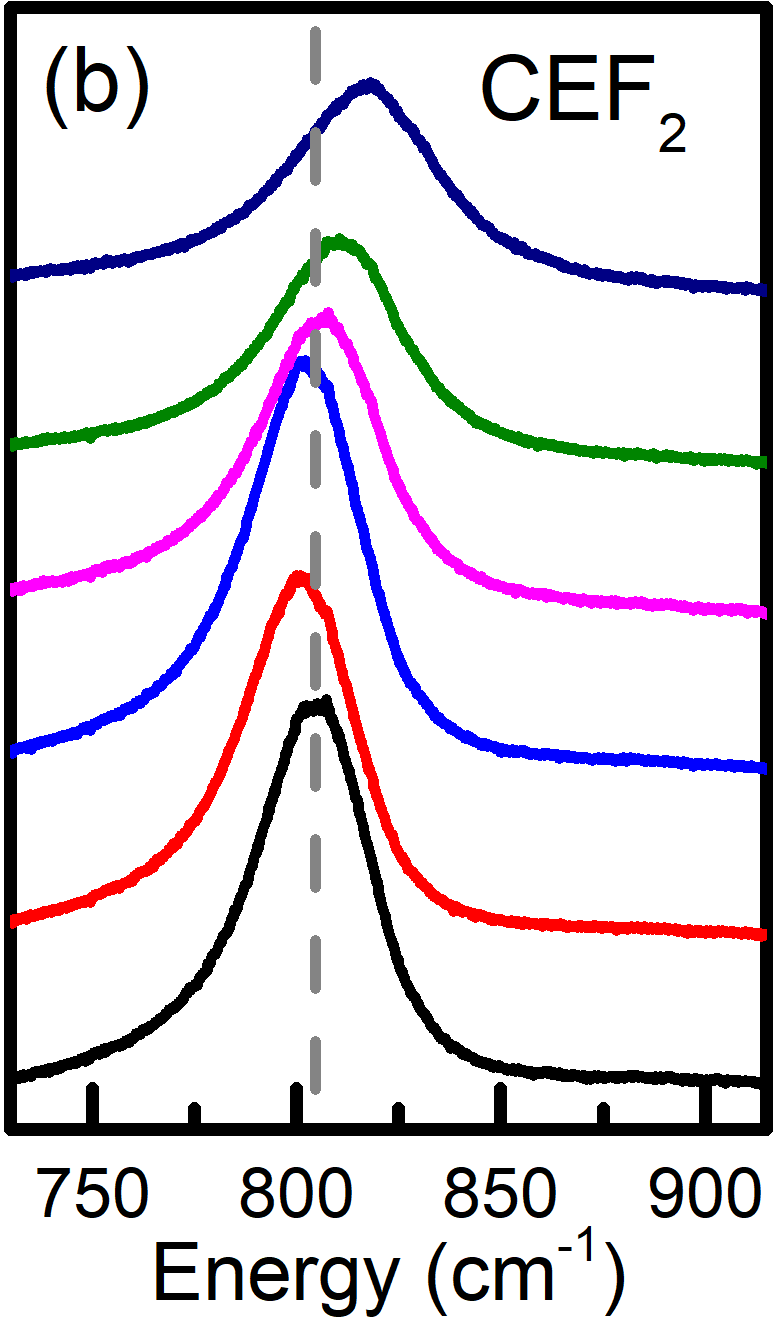} \label{Fig2b}} \\
\subfloat{\includegraphics{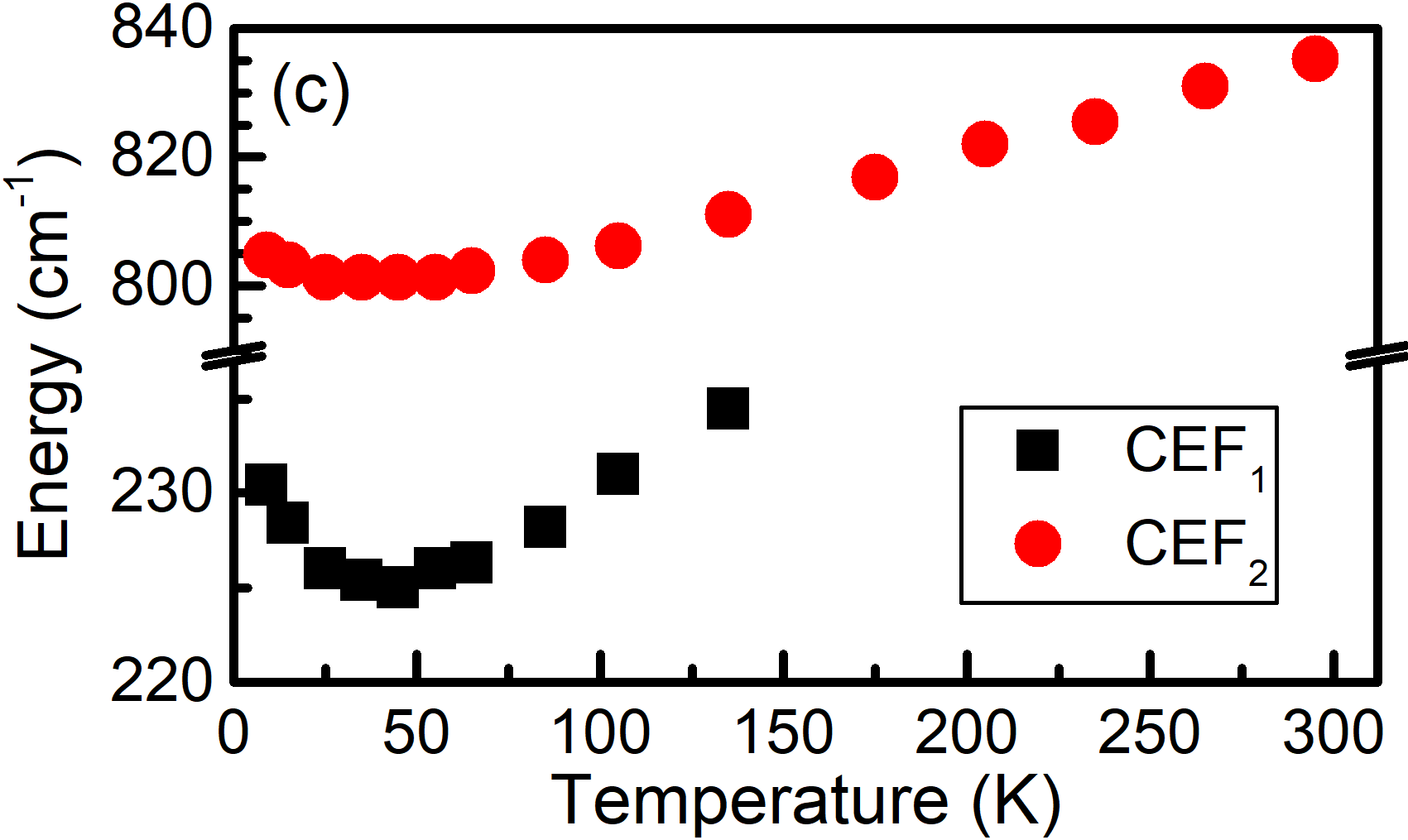} \label{Fig2c}}
\caption{\label{Fig.2}(a),(b) Temperature dependences of Raman intensities of CEF excitations in \ceri\,for $T > T_\text{N}$. The dashed lines are a guide to the eye showing the shifts in energies of the two peaks with temperature. (c) Summary of temperature dependences of CEF excitation energies.}
\end{figure}

\textit{Crystal Electric Field (CEF) excitations for $T > T_\text{N}$} -- In addition to the phonons, two broad peaks exhibiting strongly temperature dependent energies and intensities are observed at low temperatures, as shown in Figs.~\ref{Fig2a} and \ref{Fig2b}. With increasing temperature, these modes decrease in intensity, which is consistent with the temperature dependences expected for CEF excitations. Indeed, the energies of these two peaks correspond well with the CEF excitations calculated for the lower-lying CEF transitions within crystal-field-split $J=\frac{5}{2}$ manifold of states in \ceri~\cite{Kol-2018, Gru-2002}. To our knowledge, the excitations at $\omega\approx230\,\text{cm}^{-1}\,\text{and}\,\,805\,\text{cm}^{-1}$ (at $T=9$~K) shown in Fig.~\ref{Fig.2} are the first reported experimental observation of CEF excitations in \ceri. \par
Notably, the CEF excitation energies exhibit anomalous temperature dependences, initially decreasing with decreasing temperature from $295$~K to $50$~K, but then increasing in energy with decreasing temperature below $50$ K. This anomalous temperature dependence is suggestive of strong electron-phonon coupling involving the CEF excitations in \ceri.

\begin{figure}
\subfloat{\includegraphics{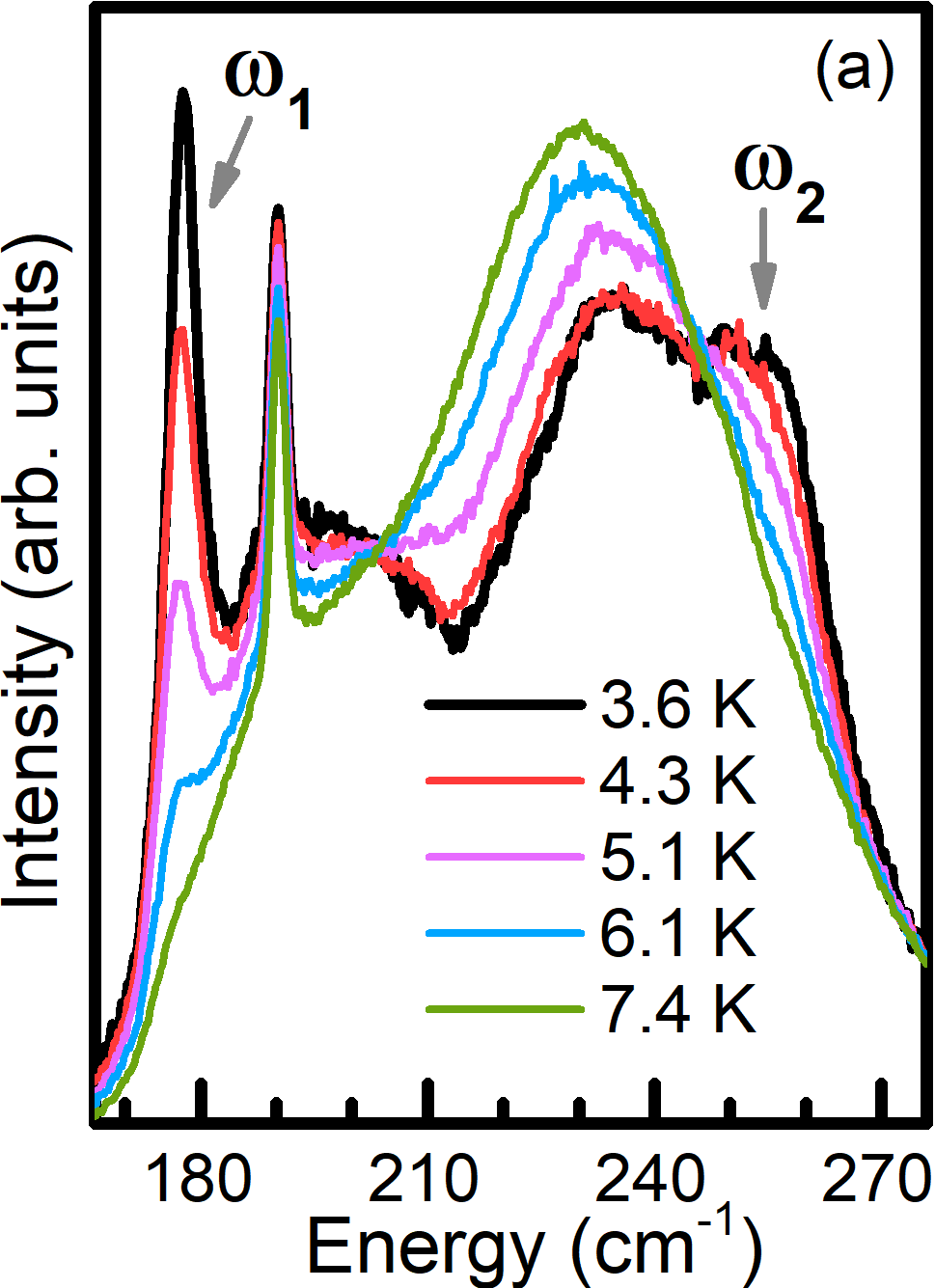} \label{Fig3a}}
\subfloat{\includegraphics{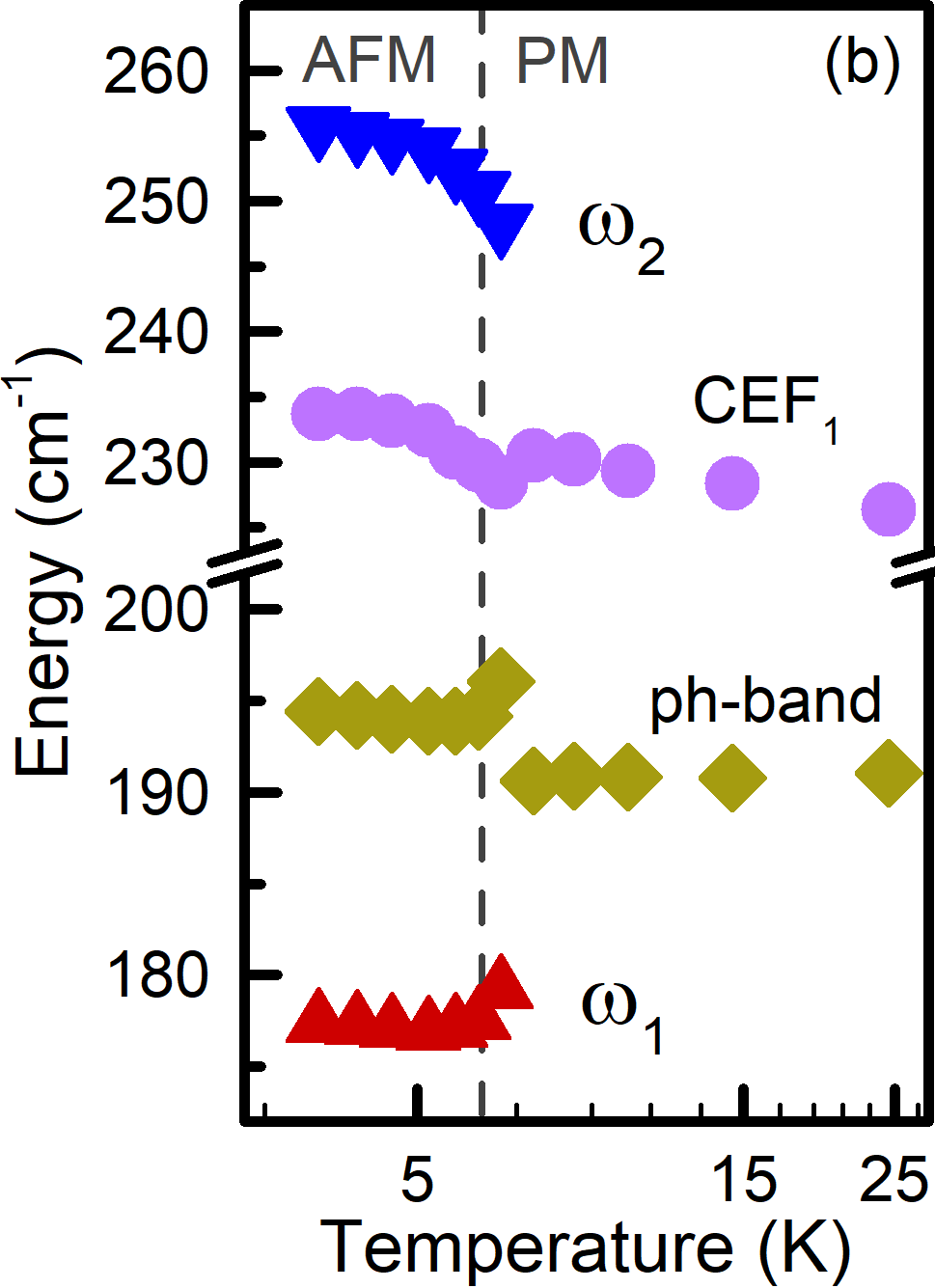} \label{Fig3b}}\\
\subfloat{\includegraphics{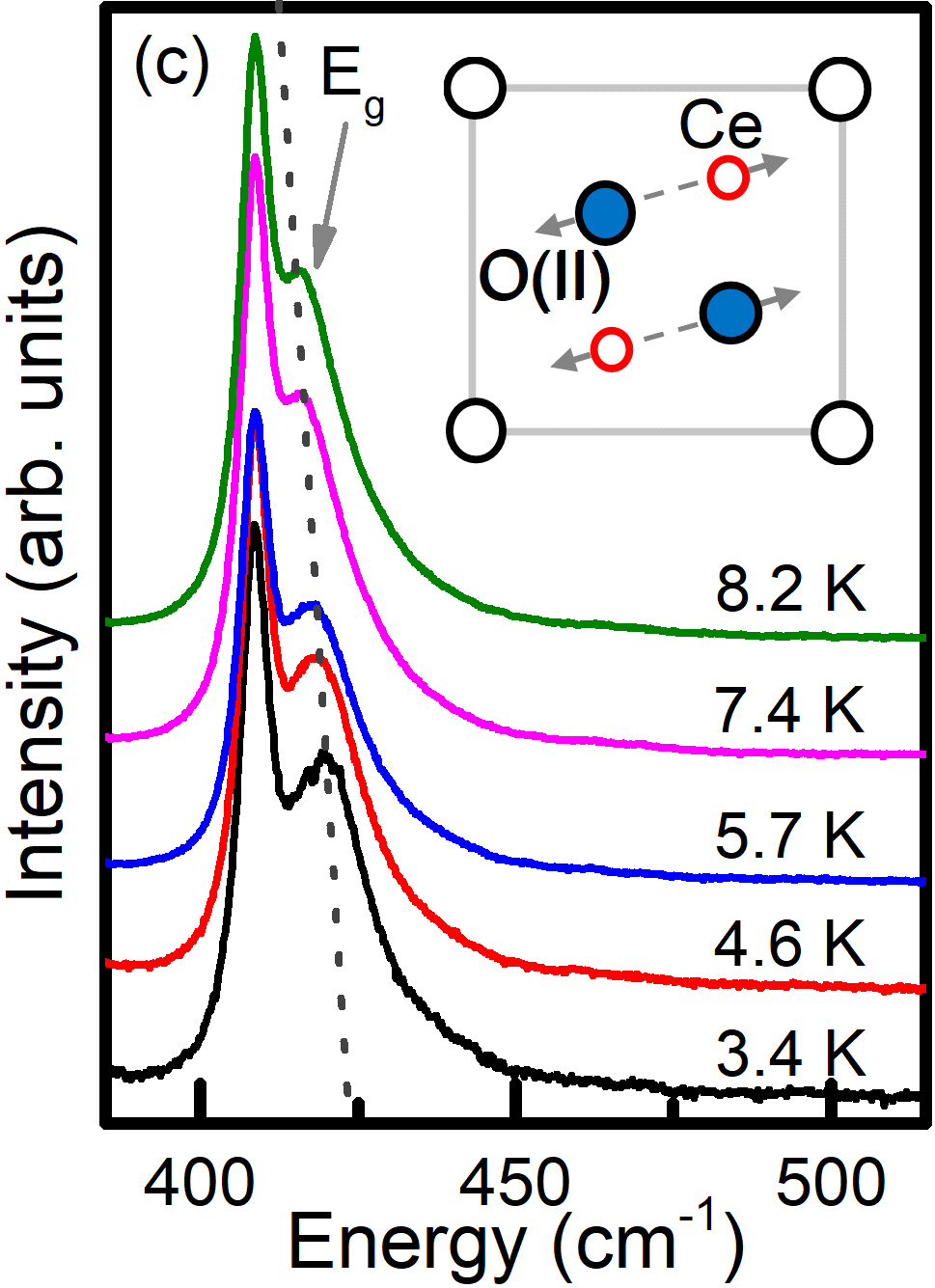} \label{Fig3c}}
\subfloat{\includegraphics{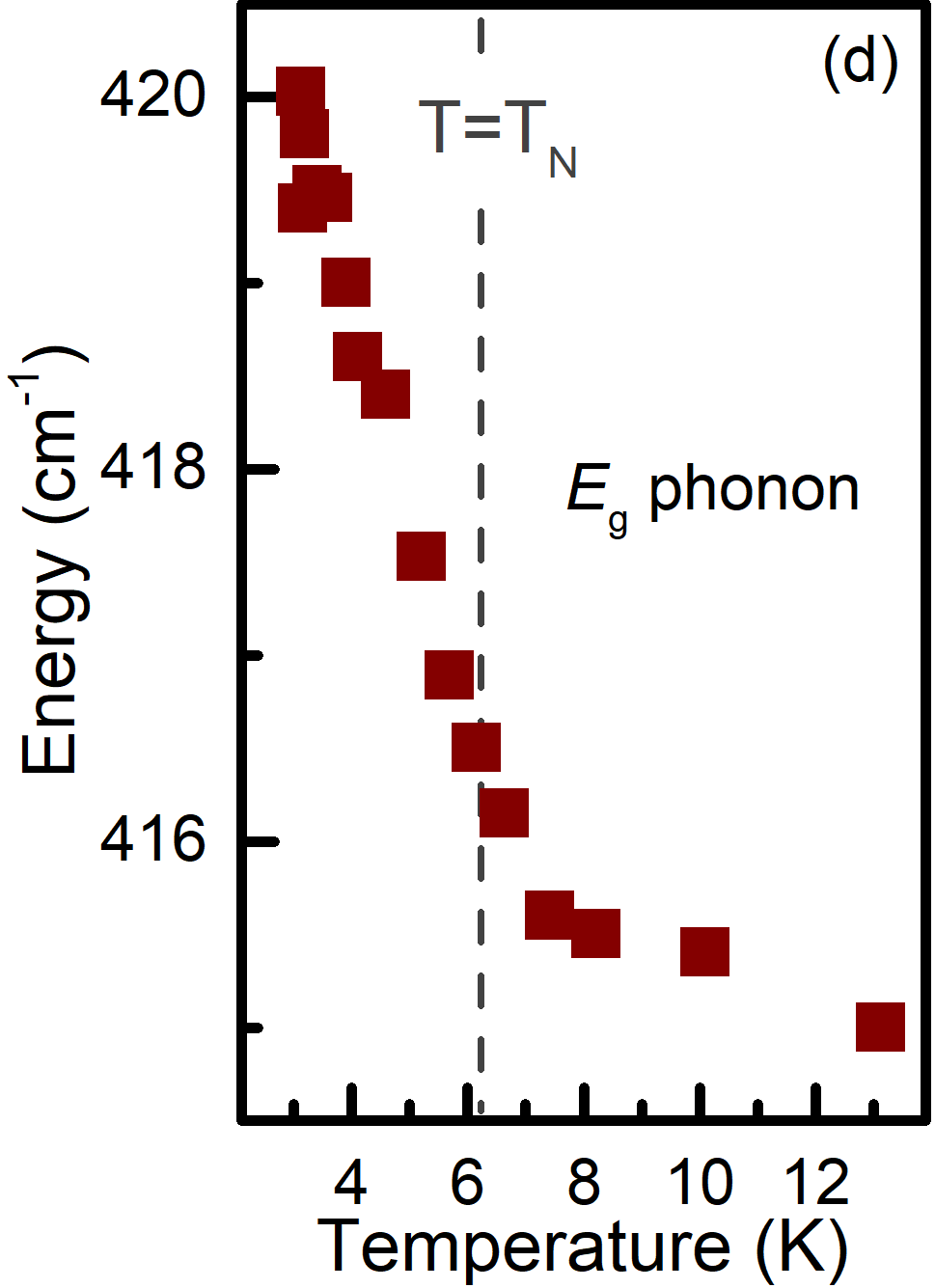} \label{Fig3d}}

\caption{\label{Fig3}(a) Raman spectrum of vibronic modes, $\omega_1$ and $\omega_2$, as a function of temperature. (b) Summary of (log scale) temperature dependences of energies of $\omega_1$, $\omega_2$, CEF$_1$ and ph-band. Symbol size used is larger than the fitting errors in peak positions. (c) Temperature dependence of Raman spectrum of the stretching $E_g$ phonon across $T_\text{N}$. Dotted line is a guide to the eye. Inset: Schematic of $E_g$ symmetry stretching vibration of Ce-O(II) bond, adapted from Refs.~\cite{Avi-2016, Gou-1981, Zar-1979}. Small empty circles denote Ce ion and large shaded circles denote O(II) ions. The large empty circles at the corners denote O(I) ions which do no participate in Raman active vibrations. Vertical axis is the c$_3$ axis. (d) Summary of dependence of $E_g$ phonon energy on temperature across $T_\text{N}$. Dashed line in (b) and (d) denotes $T=T_\text{N}$ boundary between the anti-ferromagnetic (AFM) and paramagnetic (PM) phases.}
\end{figure}

\textit{Vibronic modes and phonon anomalies for $T<T_\text{N}$} --  In the N\'{e}el state, two new modes develop (see Fig.~\ref{Fig3a}): a sharp mode at $\omega_1\approx177\,\text{cm}^{-1}$ and a broader mode at $\omega_2\approx250\,\text{cm}^{-1}$. Below $T_\text{N}$, the mode at $\omega_2$ gains intensity while the 230 cm$^{-1}$ CEF$_1$ excitation loses intensity, suggesting that there is a transfer of spectral weight from CEF$_1$ to the $\omega_2$ mode with decreasing temperature through $T_\text{N}$. The energies of the modes at $\omega_1$ and $\omega_2$ are roughly an order-of-magnitude too large for zone-center magnon or electromagnon modes given the low magnetic ordering temperature ($T_\text{N} \approx 6.2\,\text{K}$) of \ceri. In particular, other magnetic Ce-based compounds with comparable N\'{e}el temperatures have magnon energies in the range 8-35 cm$^{-1}$~\cite{Pie-1984, Osb-1987, Sch-2003, Hil-2012, Smi-2013}, well below the energies of the two new modes observed in \ceri. Consequently, we rule out a magnetic origin for the new excitations observed below $T_\text{N}$ in \ceri.

On the other hand, the two emergent modes in the magnetodielectric N\'{e}el state of \ceri\,have all the characteristics of vibronic excitations, i.e., coupled vibrational and f-electron quantum modes. Thalmeier and Fulde first predicted that strong electron-phonon coupling between a crystal field level and an energetically proximate phonon band can lead to two "vibronic" states having mixed electron-phonon character~\cite{Tha-1982}; such vibronic modes have been observed in many rare-earth compounds with atomic-like crystal field levels in the phonon energy region, including CeAl$_2$~\cite{Tha-1982}, CeCu$_2$~\cite{Loe-2003}, Ho$_2$Ti$_2$O$_7$~\cite{Gau-2018}, CeCuAl$_3$~\cite{Adj-2012} and NdBa$_2$Cu$_3$O$_7$~\cite{Hey-1991}. Several characteristics identify the two new modes observed below $T_\text{N}$ in \ceri\,as vibronic modes involving the CEF$_1$ level and a phonon band: First, the development of two new modes, $\omega_1$ and $\omega_2$, in the same energy region as the CEF$_1$ and phonon modes is consistent with the vibronic mode interpretation, as shown more quantitatively below. Second, the increasing separation between the $\omega_1$ and $\omega_2$ mode energies with decreasing temperature (see Fig.~\ref{Fig3a} and ~\ref{Fig3b}) are consistent with an increasing level repulsion with decreasing temperature within the Thalmeier-Fulde vibronic model. Finally, the transfer of scattering strength from the electronic CEF$_1$ excitation to the broader vibronic mode is strong evidence for mixed electronic character associated with the $\omega_2$ mode.

In addition to the emergent vibronic modes, a broad feature at $\omega\approx194\,\text{cm}^{-1}$ develops at low temperatures, especially below $7$ K (for clarity see Fig. S1 in Supplemental Material). As the four expected single-phonon excitations and the two CEF excitations in this energy range have already been accounted for, this broad feature likely involves a 2-phonon band that is involved in the formation of the vibronic modes. Figure~\ref{Fig3b} summarizes the temperature dependences of the peak energies of the two vibronic modes, the CEF$_1$ excitation and the 2-phonon band, extracted from curve fits to the Raman spectra. The curve fitting procedure is described in the Supplemental Material (see Fig. S1).\par
The Thalmeier-Fulde description predicts that the relative mixing of electron-phonon character associated with the vibronic modes depends on the energy difference between the coupled phonon band and the CEF level~\cite{Tha-1982}. In \ceri, the appearance of a relatively narrow "phonon-like" mode at $\omega_1\approx177\,\text{cm}^{-1}$ and a broader "electron-like" mode at $\omega_2\approx250\,\text{cm}^{-1}$ is consistent with the relatively large separation between the coupled phonon band at $194\,\text{cm}^{-1}$ and the CEF$_1$ level.

Unlike previously observed vibronic modes in rare-earth materials~\cite{Tha-1982, Loe-2003,Gau-2018,Adj-2012,Hey-1991}, the vibronic modes observed in \ceri\,are noteworthy in that they emerge in the MD regime below $T_\text{N}$. Importantly, the integrated Raman scattering intensity of an excitation is proportional to the modulation of the dielectric response due to that excitation~\cite{Hay-Lou, Lus-2004, Ash-2011}. The significant increase in the intensity of both the vibronic modes observed below $T_\text{N}$ (see Fig.~\ref{Fig3a}) is consistent with an enhanced modulation of the dielectric susceptibility due to these modes, as expected in the MD regime in the N\'{e}el state of \ceri. In particular, the increased coupling between the CEF electronic and phononic modes below $T_\text{N}$ is expected to result in enhanced fluctuations in the electronic levels, which is known to cause increased fluctuations of the dielectric response~\cite{Wem-1969}.

To extract more quantitative information from the emergent $\omega_1$ and $\omega_2$ modes in \ceri, we apply the Thalmeier-Fulde model for vibronic mode energies resulting from a coupling between a CEF excitation and a phonon band ~\cite{Tha-1982}:

\begin{equation}
\omega_{1,2} = \frac{\omega_{CEF}+\omega_{ph}}{2}  \mp \sqrt{\Big(\frac{\omega_{CEF} - \omega_{ph}}{2}\Big)^2+V^2}
\label{eq1}
\end{equation}

where $\omega_{CEF}$ and $\omega_{ph}$ denote the energies of the CEF excitation and the phonon band, respectively, and $V$ denotes the electron-phonon coupling strength between them. Using the measured vibronic energies, $\omega_1$ and $\omega_2$, and the involved CEF$_1$ excitation energy in Eq.~\eqref{eq1}, we can extract the temperature dependence of $V$ and $\omega_{ph}$ (see Table II in the Supplemental Material). The calculated phonon band energy ($\approx 199\,\text{cm}^{-1}$) is close to the experimentally observed 2-phonon band centered at $\omega_{ph} \approx 194\,\text{cm}^{-1}$, consistent with our assumption that this band is involved in the vibronic coupling.

\begin{figure}
\subfloat{\includegraphics{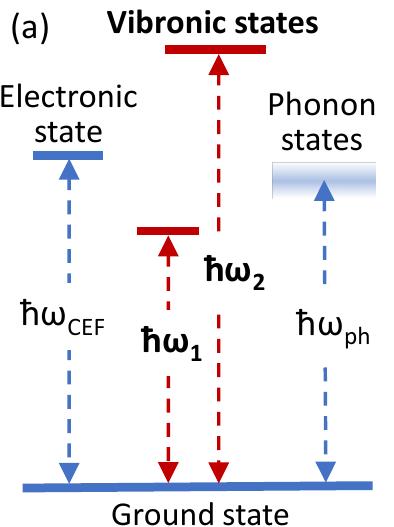} \label{Fig4a}}
\subfloat{\includegraphics{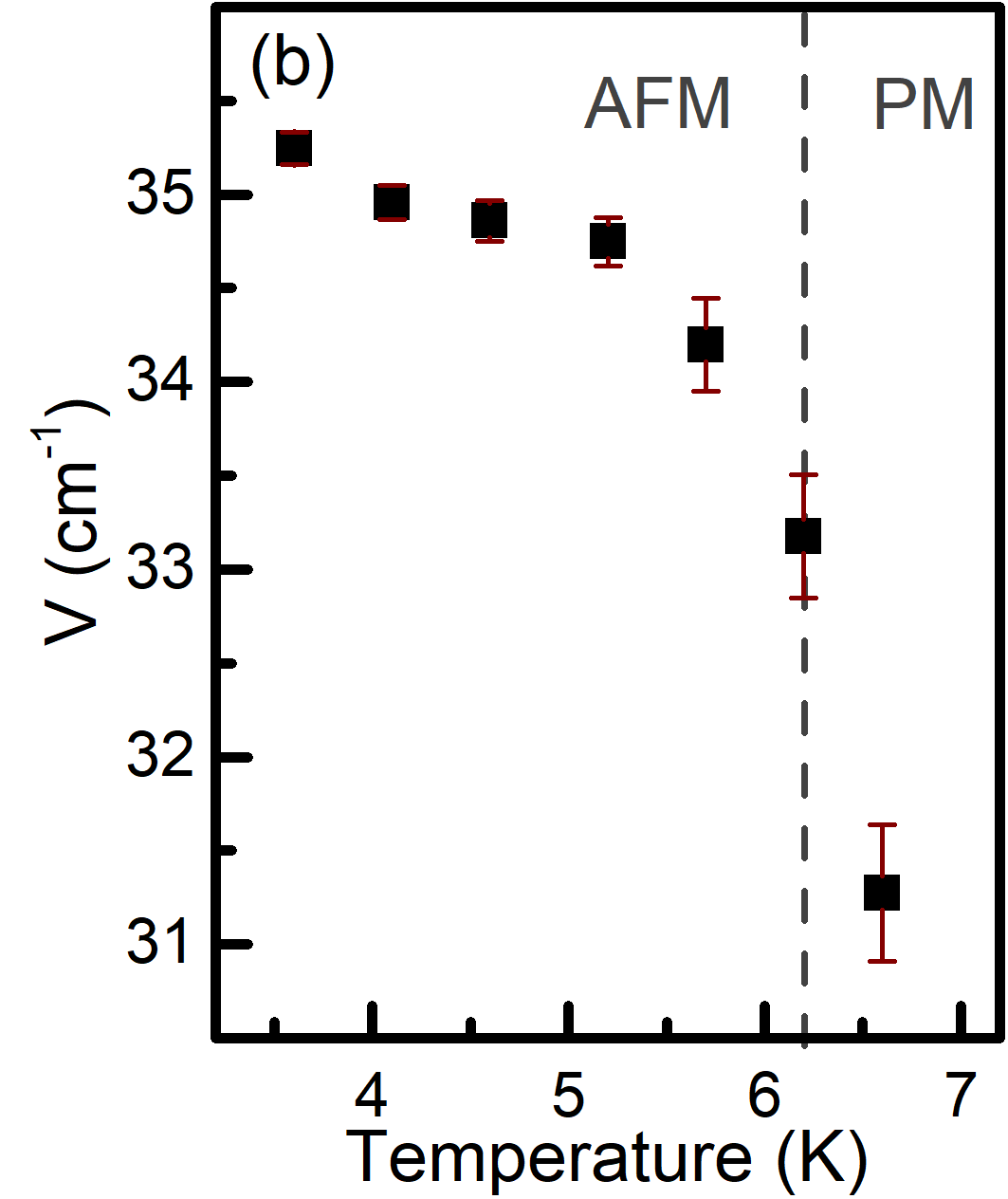} \label{Fig4b}}
\caption{\label{Fig4}(a) Schematic representation of the coupling of electronic (CEF) and vibrational states to form new bound states called "vibronic states". (b) Temperature dependence of the electron-phonon coupling constant, $V$ (extracted from Eq.~\eqref{eq1}). Dashed line denotes $T=T_\text{N}$. The fitting errors in peak positions were used to estimate the error bars in $V$ as per Eq.~\eqref{eq1}}
\end{figure}

The resulting temperature dependence of the electron-phonon coupling strength, $V$, is plotted in Fig.~\ref{Fig4b}, showing that the emergence of the vibronic modes below $T_\text{N}$ in \ceri\,results from a rapid increase in electron-phonon coupling strength through the N\'{e}el transition. Significantly, the increase in electron-phonon coupling shown in Fig.~\ref{Fig4b} coincides with the abrupt increase in the 416 cm$^{-1}$ $E_g$ - symmetry stretching vibration below $T_\text{N}$ (see Figs.~\ref{Fig3c} and \ref{Fig3d}), which suggests a decrease in the Ce-O(II) bond length involved in this vibration. Consequently, we propose that the increasing vibronic mode coupling parameter and resultant emergence of vibronic modes in \ceri\,is caused by magnetostructural changes associated with the N\'{e}el ordering, which result in an abrupt shortening of the Ce-O(II) bond below $T_\text{N}$.

\begin{figure}
\subfloat{\includegraphics{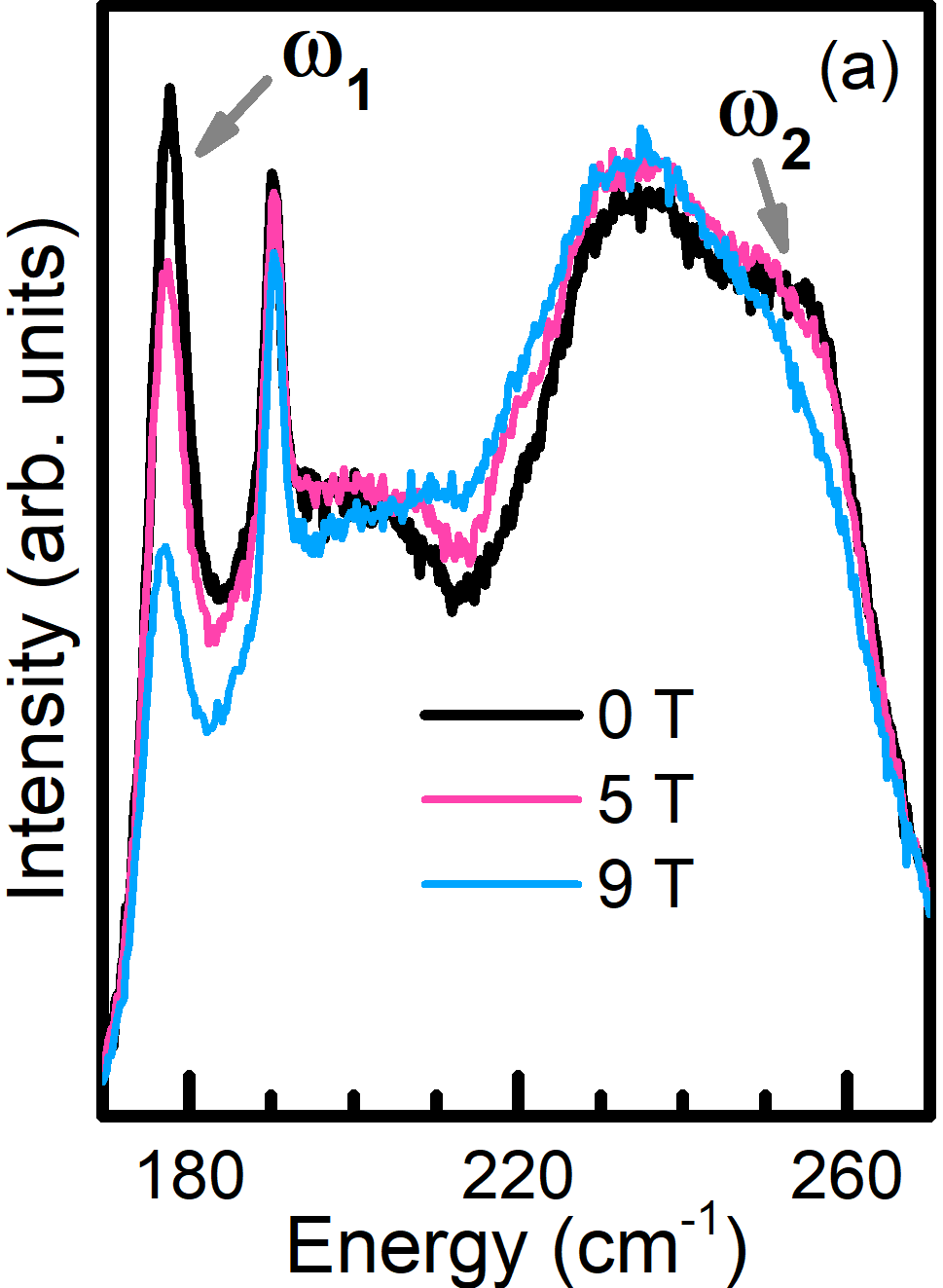} \label{Fig5a}}
\subfloat{\includegraphics{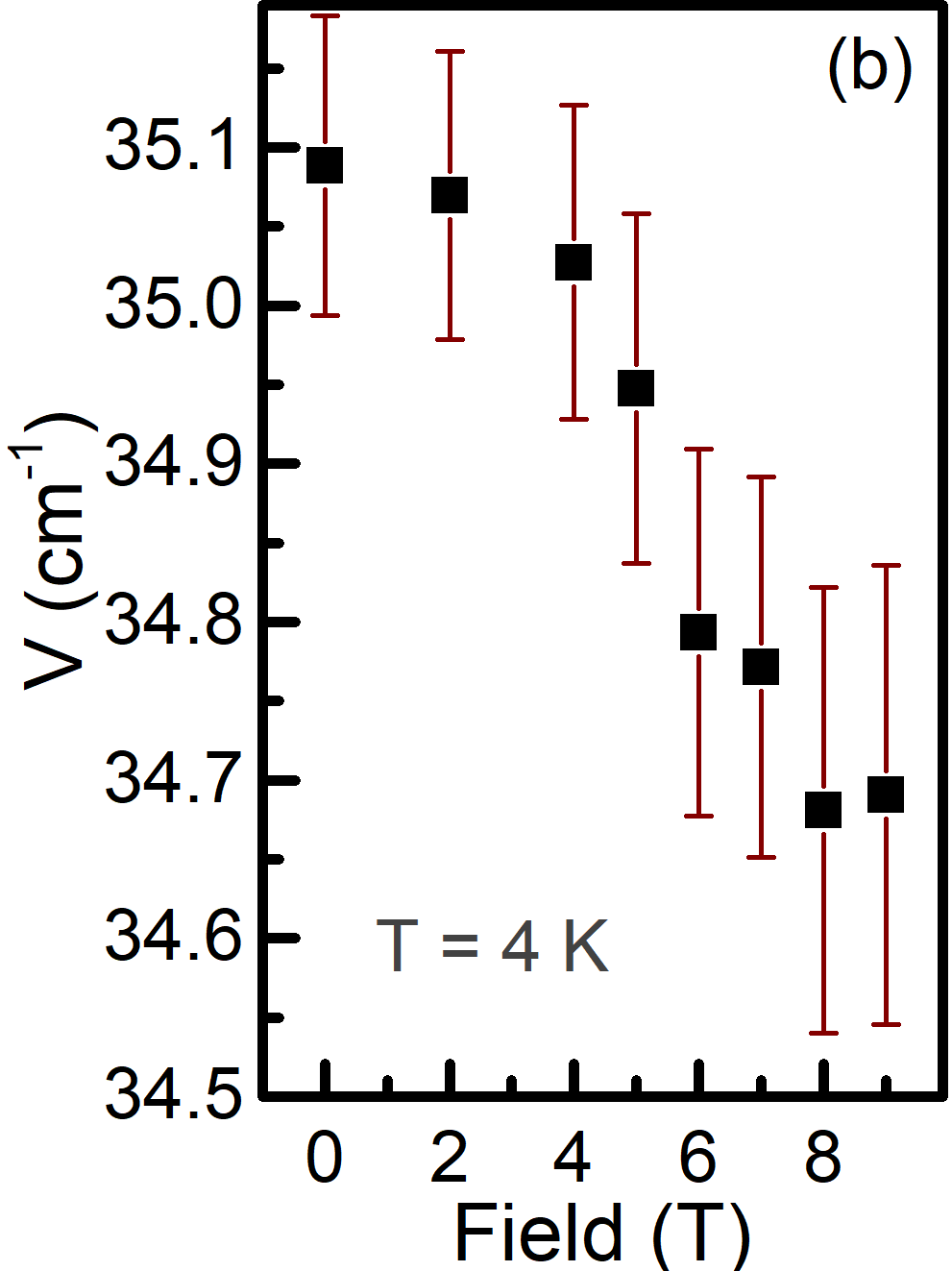} \label{Fig5b}}
\caption{\label{Fig5}(a) Raman spectrum of vibronic modes as a function of magnetic field at $T = 4$ K. (b) Summary of field dependence of electron-phonon coupling constant, $V$ (extracted from Eq.~\eqref{eq1}). The fitting errors in peak positions were used to estimate the error bars in $V$ as per Eq.~\eqref{eq1}}
\end{figure}

To test the above hypothesis, we conducted magnetic field-dependent Raman measurements to study the effects of magnetic field on the vibronic and phonon modes. Figure~\ref{Fig5a} shows that an increasing magnetic field leads to a decrease of the vibronic mode intensities and a slight reduction of the level repulsion between the $\omega_1$ and $\omega_2$ modes, primarily resulting from a shift of the $\omega_2$ mode to lower energies. Analyzing the field-dependent results using Eq.~\eqref{eq1}, the slight decrease in the energy separation between $\omega_1$ and $\omega_2$ vibronic modes with increasing field is consistent with a small decrease in the vibronic mode coupling with applied field, as summarized in Fig.~\ref{Fig5b}. Consistent with this result, the 416 cm$^{-1}$ $E_g$ - phonon exhibits a decrease in energy with increasing field for $T<T_\text{N}$, indicating that the associated Ce-O(II) bond lengthens slightly with increasing magnetic field in the N\'{e}el state. These results support the hypothesis that magnetostructural changes below $T_\text{N}$ are responsible for the shortening of the Ce-O(II) bond and the emergence of vibronic modes in the MD regime of \ceri. In this regard, the emergence of vibronic modes in the MD phase of \ceri\,is analogous to the emergence of electromagnons that develop via magnetostriction in the magnetoelectric phase of magnetoelectric materials~\cite{Agu-2009, Sus-2007}.\par

Additionally, the decrease of the vibronic mode intensities with applied field is consistent with a reduced modulation of the dielectric response by these modes in the N\'{e}el state. Altogether, these field-dependent results provide evidence that the field-tunable coupling between the CEF and phonon excitations is correlated with the unusual MD behavior in \ceri below $T_\text{N}$, presumably because of enhanced fluctuations of the CEF levels by phonons~\cite{Wem-1969}.

\begin{figure}
\subfloat{\includegraphics{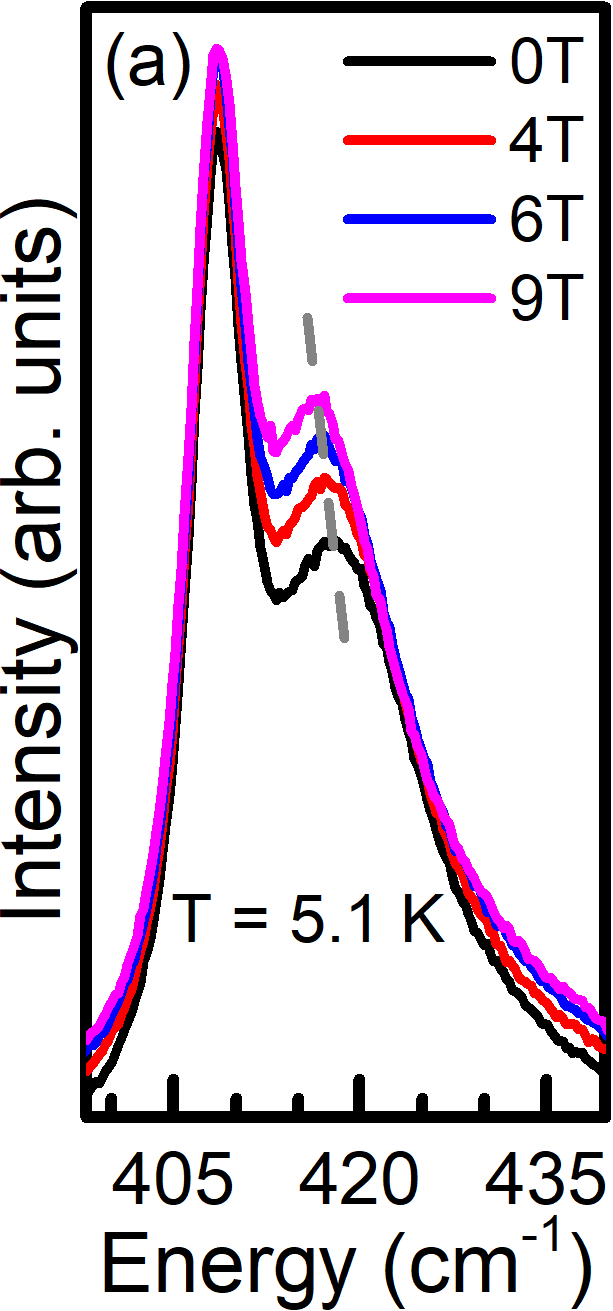} \label{Fig6a}}
\subfloat{\includegraphics{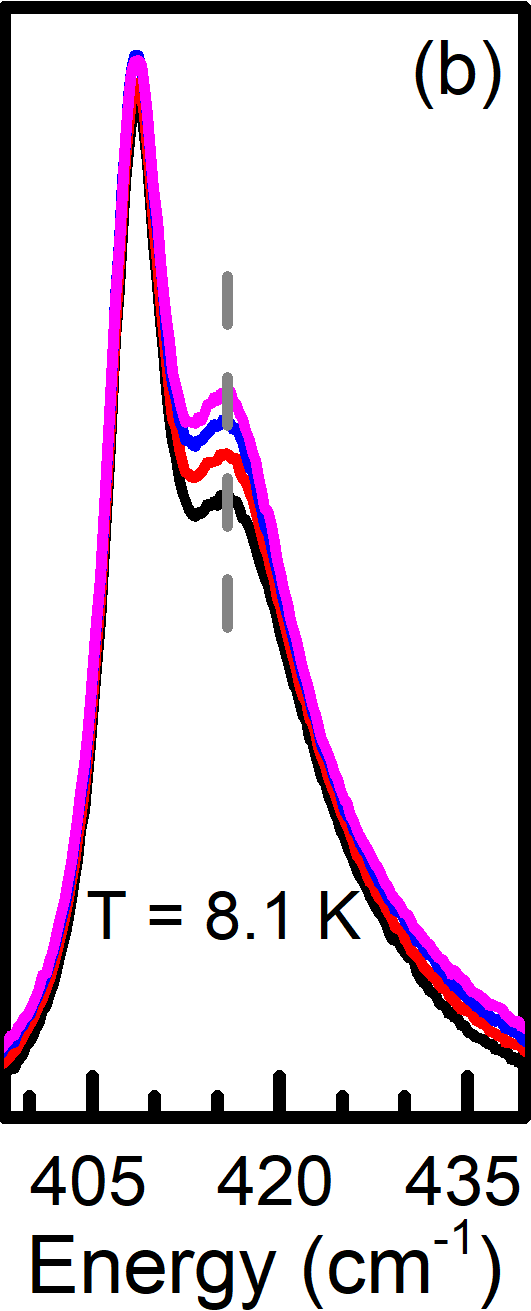} \label{Fig6b}}
\subfloat{\includegraphics{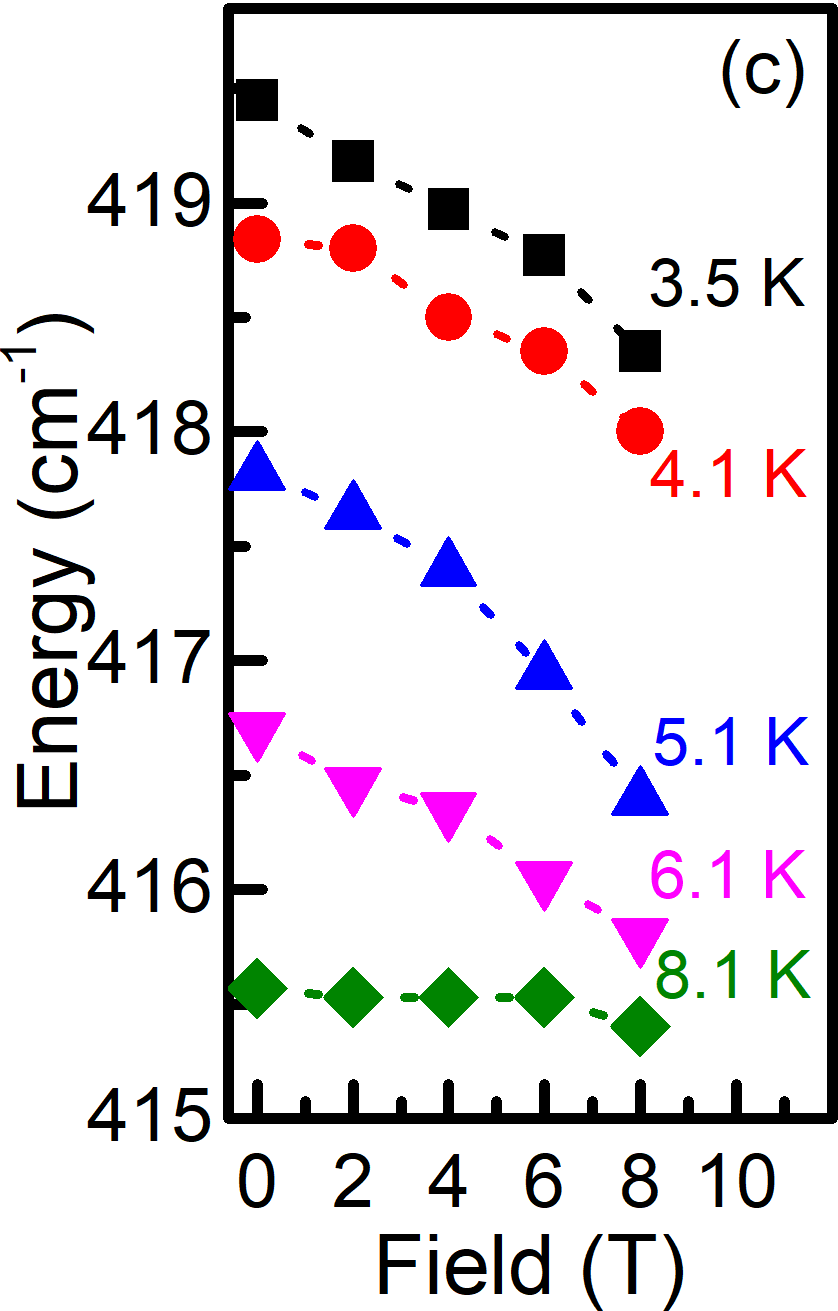} \label{Fig6c}}
\caption{\label{Fig6}Raman spectrum of stretching $E_g$ phonon as a function of field in (a) magnetic phase, $T=5.1$\,K and (b) paramagnetic phase, $T=8.1$\,K. (c) Summary of $E_g$ phonon energy as a function of magnetic field at various temperatures. Dotted lines are guide to the eye.}
\end{figure}

\textit{Conclusion}-- The localized f-orbitals in \ceri\,result in an energetic overlap between the CEF and phonon excitations, which is favorable for strong electron-phonon (vibronic) coupling. We provide evidence that magnetostructural changes below $T_\text{N} (\approx 6.2\,\text{K})$ in \ceri\,lead to a rapid enhancement in CEF-phonon coupling, which is manifest in the emergence of two coupled vibronic modes. Our results suggest that the strong interactions between the CEF electronic and phononic subsystems lead to increased fluctuations of the dielectric response below $T_\text{N}$, which we argue result from increased fluctuations of the CEF levels. The field-tunable energies and intensities of the vibronic modes suggest that the coupling between the CEF and phonon states--and consequently the degree to which the dielectric response is modulated by phonons--is reduced with increasing field. These results suggest a distinct mechanism for magnetodielectricity in rare-earth materials like \ceri, which is not expected in TM oxides because of the significantly higher CEF energies associated with d-orbital materials.

\textit{Acknowledgement}-- Research was supported by the National Science Foundation under Grant NSF DMR 1800982. T.K. was supported by NIMS internal projects PA5160 and PA4020.


\begin{thebibliography}{28}%
\makeatletter
\providecommand \@ifxundefined [1]{%
 \@ifx{#1\undefined}
}%
\providecommand \@ifnum [1]{%
 \ifnum #1\expandafter \@firstoftwo
 \else \expandafter \@secondoftwo
 \fi
}%
\providecommand \@ifx [1]{%
 \ifx #1\expandafter \@firstoftwo
 \else \expandafter \@secondoftwo
 \fi
}%
\providecommand \natexlab [1]{#1}%
\providecommand \enquote  [1]{``#1''}%
\providecommand \bibnamefont  [1]{#1}%
\providecommand \bibfnamefont [1]{#1}%
\providecommand \citenamefont [1]{#1}%
\providecommand \href@noop [0]{\@secondoftwo}%
\providecommand \href [0]{\begingroup \@sanitize@url \@href}%
\providecommand \@href[1]{\@@startlink{#1}\@@href}%
\providecommand \@@href[1]{\endgroup#1\@@endlink}%
\providecommand \@sanitize@url [0]{\catcode `\\12\catcode `\$12\catcode
  `\&12\catcode `\#12\catcode `\^12\catcode `\_12\catcode `\%12\relax}%
\providecommand \@@startlink[1]{}%
\providecommand \@@endlink[0]{}%
\providecommand \url  [0]{\begingroup\@sanitize@url \@url }%
\providecommand \@url [1]{\endgroup\@href {#1}{\urlprefix }}%
\providecommand \urlprefix  [0]{URL }%
\providecommand \Eprint [0]{\href }%
\providecommand \doibase [0]{http://dx.doi.org/}%
\providecommand \selectlanguage [0]{\@gobble}%
\providecommand \bibinfo  [0]{\@secondoftwo}%
\providecommand \bibfield  [0]{\@secondoftwo}%
\providecommand \translation [1]{[#1]}%
\providecommand \BibitemOpen [0]{}%
\providecommand \bibitemStop [0]{}%
\providecommand \bibitemNoStop [0]{.\EOS\space}%
\providecommand \EOS [0]{\spacefactor3000\relax}%
\providecommand \BibitemShut  [1]{\csname bibitem#1\endcsname}%
\let\auto@bib@innerbib\@empty
\bibitem [{\citenamefont {Cheong}\ and\ \citenamefont
  {Mostovoy}(2007)}]{Che-2007}%
  \BibitemOpen
  \bibfield  {author} {\bibinfo {author} {\bibfnamefont {S.-W.}\ \bibnamefont
  {Cheong}}\ and\ \bibinfo {author} {\bibfnamefont {M.}~\bibnamefont
  {Mostovoy}},\ }\href@noop {} {\bibfield  {journal} {\bibinfo  {journal}
  {Nature Materials}\ }\textbf {\bibinfo {volume} {6}},\ \bibinfo {pages} {13
  EP } (\bibinfo {year} {2007})}\BibitemShut {NoStop}%
\bibitem [{\citenamefont {Mufti}\ \emph {et~al.}(2010)\citenamefont {Mufti},
  \citenamefont {Nugroho}, \citenamefont {Blake},\ and\ \citenamefont
  {Palstra}}]{Muf-2010}%
  \BibitemOpen
  \bibfield  {author} {\bibinfo {author} {\bibfnamefont {N.}~\bibnamefont
  {Mufti}}, \bibinfo {author} {\bibfnamefont {A.~A.}\ \bibnamefont {Nugroho}},
  \bibinfo {author} {\bibfnamefont {G.~R.}\ \bibnamefont {Blake}}, \ and\
  \bibinfo {author} {\bibfnamefont {T.~T.~M.}\ \bibnamefont {Palstra}},\ }\href
  {http://stacks.iop.org/0953-8984/22/i=7/a=075902} {\bibfield  {journal}
  {\bibinfo  {journal} {Journal of Physics: Condensed Matter}\ }\textbf
  {\bibinfo {volume} {22}},\ \bibinfo {pages} {075902} (\bibinfo {year}
  {2010})}\BibitemShut {NoStop}%
\bibitem [{\citenamefont {Vald\'es~Aguilar}\ \emph {et~al.}(2009)\citenamefont
  {Vald\'es~Aguilar}, \citenamefont {Mostovoy}, \citenamefont {Sushkov},
  \citenamefont {Zhang}, \citenamefont {Choi}, \citenamefont {Cheong},\ and\
  \citenamefont {Drew}}]{Agu-2009}%
  \BibitemOpen
  \bibfield  {author} {\bibinfo {author} {\bibfnamefont {R.}~\bibnamefont
  {Vald\'es~Aguilar}}, \bibinfo {author} {\bibfnamefont {M.}~\bibnamefont
  {Mostovoy}}, \bibinfo {author} {\bibfnamefont {A.~B.}\ \bibnamefont
  {Sushkov}}, \bibinfo {author} {\bibfnamefont {C.~L.}\ \bibnamefont {Zhang}},
  \bibinfo {author} {\bibfnamefont {Y.~J.}\ \bibnamefont {Choi}}, \bibinfo
  {author} {\bibfnamefont {S.-W.}\ \bibnamefont {Cheong}}, \ and\ \bibinfo
  {author} {\bibfnamefont {H.~D.}\ \bibnamefont {Drew}},\ }\href {\doibase
  10.1103/PhysRevLett.102.047203} {\bibfield  {journal} {\bibinfo  {journal}
  {Phys. Rev. Lett.}\ }\textbf {\bibinfo {volume} {102}},\ \bibinfo {pages}
  {047203} (\bibinfo {year} {2009})}\BibitemShut {NoStop}%
\bibitem [{\citenamefont {Sushkov}\ \emph {et~al.}(2007)\citenamefont
  {Sushkov}, \citenamefont {Aguilar}, \citenamefont {Park}, \citenamefont
  {Cheong},\ and\ \citenamefont {Drew}}]{Sus-2007}%
  \BibitemOpen
  \bibfield  {author} {\bibinfo {author} {\bibfnamefont {A.~B.}\ \bibnamefont
  {Sushkov}}, \bibinfo {author} {\bibfnamefont {R.~V.}\ \bibnamefont
  {Aguilar}}, \bibinfo {author} {\bibfnamefont {S.}~\bibnamefont {Park}},
  \bibinfo {author} {\bibfnamefont {S.-W.}\ \bibnamefont {Cheong}}, \ and\
  \bibinfo {author} {\bibfnamefont {H.~D.}\ \bibnamefont {Drew}},\ }\href
  {\doibase 10.1103/PhysRevLett.98.027202} {\bibfield  {journal} {\bibinfo
  {journal} {Phys. Rev. Lett.}\ }\textbf {\bibinfo {volume} {98}},\ \bibinfo
  {pages} {027202} (\bibinfo {year} {2007})}\BibitemShut {NoStop}%
\bibitem [{\citenamefont {Pimenov}\ \emph {et~al.}(2006)\citenamefont
  {Pimenov}, \citenamefont {Mukhin}, \citenamefont {Ivanov}, \citenamefont
  {Travkin}, \citenamefont {Balbashov},\ and\ \citenamefont
  {Loidl}}]{Pim-2006}%
  \BibitemOpen
  \bibfield  {author} {\bibinfo {author} {\bibfnamefont {A.}~\bibnamefont
  {Pimenov}}, \bibinfo {author} {\bibfnamefont {A.~A.}\ \bibnamefont {Mukhin}},
  \bibinfo {author} {\bibfnamefont {V.~Y.}\ \bibnamefont {Ivanov}}, \bibinfo
  {author} {\bibfnamefont {V.~D.}\ \bibnamefont {Travkin}}, \bibinfo {author}
  {\bibfnamefont {A.~M.}\ \bibnamefont {Balbashov}}, \ and\ \bibinfo {author}
  {\bibfnamefont {A.}~\bibnamefont {Loidl}},\ }\href
  {https://doi.org/10.1038/nphys212} {\bibfield  {journal} {\bibinfo  {journal}
  {Nature Physics}\ }\textbf {\bibinfo {volume} {2}},\ \bibinfo {pages} {97 EP
  } (\bibinfo {year} {2006})}\BibitemShut {NoStop}%
\bibitem [{\citenamefont {Dong}\ \emph {et~al.}(2015)\citenamefont {Dong},
  \citenamefont {Liu}, \citenamefont {Cheong},\ and\ \citenamefont
  {Ren}}]{Don-2015}%
  \BibitemOpen
  \bibfield  {author} {\bibinfo {author} {\bibfnamefont {S.}~\bibnamefont
  {Dong}}, \bibinfo {author} {\bibfnamefont {J.-M.}\ \bibnamefont {Liu}},
  \bibinfo {author} {\bibfnamefont {S.-W.}\ \bibnamefont {Cheong}}, \ and\
  \bibinfo {author} {\bibfnamefont {Z.}~\bibnamefont {Ren}},\ }\href {\doibase
  10.1080/00018732.2015.1114338} {\bibfield  {journal} {\bibinfo  {journal}
  {Advances in Physics}\ }\textbf {\bibinfo {volume} {64}},\ \bibinfo {pages}
  {519} (\bibinfo {year} {2015})},\ \Eprint
  {http://arxiv.org/abs/https://doi.org/10.1080/00018732.2015.1114338}
  {https://doi.org/10.1080/00018732.2015.1114338} \BibitemShut {NoStop}%
\bibitem [{\citenamefont {Ortega}\ \emph {et~al.}(2015)\citenamefont {Ortega},
  \citenamefont {Kumar}, \citenamefont {Scott},\ and\ \citenamefont
  {Katiyar}}]{Ort-2015}%
  \BibitemOpen
  \bibfield  {author} {\bibinfo {author} {\bibfnamefont {N.}~\bibnamefont
  {Ortega}}, \bibinfo {author} {\bibfnamefont {A.}~\bibnamefont {Kumar}},
  \bibinfo {author} {\bibfnamefont {J.~F.}\ \bibnamefont {Scott}}, \ and\
  \bibinfo {author} {\bibfnamefont {R.~S.}\ \bibnamefont {Katiyar}},\ }\href
  {http://stacks.iop.org/0953-8984/27/i=50/a=504002} {\bibfield  {journal}
  {\bibinfo  {journal} {Journal of Physics: Condensed Matter}\ }\textbf
  {\bibinfo {volume} {27}},\ \bibinfo {pages} {504002} (\bibinfo {year}
  {2015})}\BibitemShut {NoStop}%
\bibitem [{\citenamefont {Juraschek}\ \emph {et~al.}(2017)\citenamefont
  {Juraschek}, \citenamefont {Fechner}, \citenamefont {Balatsky},\ and\
  \citenamefont {Spaldin}}]{Jur-2017}%
  \BibitemOpen
  \bibfield  {author} {\bibinfo {author} {\bibfnamefont {D.~M.}\ \bibnamefont
  {Juraschek}}, \bibinfo {author} {\bibfnamefont {M.}~\bibnamefont {Fechner}},
  \bibinfo {author} {\bibfnamefont {A.~V.}\ \bibnamefont {Balatsky}}, \ and\
  \bibinfo {author} {\bibfnamefont {N.~A.}\ \bibnamefont {Spaldin}},\ }\href
  {\doibase 10.1103/PhysRevMaterials.1.014401} {\bibfield  {journal} {\bibinfo
  {journal} {Phys. Rev. Materials}\ }\textbf {\bibinfo {volume} {1}},\ \bibinfo
  {pages} {014401} (\bibinfo {year} {2017})}\BibitemShut {NoStop}%
\bibitem [{\citenamefont {Katsura}\ \emph {et~al.}(2007)\citenamefont
  {Katsura}, \citenamefont {Balatsky},\ and\ \citenamefont
  {Nagaosa}}]{Kat-2007}%
  \BibitemOpen
  \bibfield  {author} {\bibinfo {author} {\bibfnamefont {H.}~\bibnamefont
  {Katsura}}, \bibinfo {author} {\bibfnamefont {A.~V.}\ \bibnamefont
  {Balatsky}}, \ and\ \bibinfo {author} {\bibfnamefont {N.}~\bibnamefont
  {Nagaosa}},\ }\href {\doibase 10.1103/PhysRevLett.98.027203} {\bibfield
  {journal} {\bibinfo  {journal} {Phys. Rev. Lett.}\ }\textbf {\bibinfo
  {volume} {98}},\ \bibinfo {pages} {027203} (\bibinfo {year}
  {2007})}\BibitemShut {NoStop}%
\bibitem [{\citenamefont {Kolodiazhnyi}\ \emph {et~al.}(2018)\citenamefont
  {Kolodiazhnyi}, \citenamefont {Sakurai}, \citenamefont {Avdeev},
  \citenamefont {Charoonsuk}, \citenamefont {Lamonova}, \citenamefont
  {Pashkevich},\ and\ \citenamefont {Kennedy}}]{Kol-2018}%
  \BibitemOpen
  \bibfield  {author} {\bibinfo {author} {\bibfnamefont {T.}~\bibnamefont
  {Kolodiazhnyi}}, \bibinfo {author} {\bibfnamefont {H.}~\bibnamefont
  {Sakurai}}, \bibinfo {author} {\bibfnamefont {M.}~\bibnamefont {Avdeev}},
  \bibinfo {author} {\bibfnamefont {T.}~\bibnamefont {Charoonsuk}}, \bibinfo
  {author} {\bibfnamefont {K.~V.}\ \bibnamefont {Lamonova}}, \bibinfo {author}
  {\bibfnamefont {Y.~G.}\ \bibnamefont {Pashkevich}}, \ and\ \bibinfo {author}
  {\bibfnamefont {B.~J.}\ \bibnamefont {Kennedy}},\ }\href {\doibase
  10.1103/PhysRevB.98.054423} {\bibfield  {journal} {\bibinfo  {journal} {Phys.
  Rev. B}\ }\textbf {\bibinfo {volume} {98}},\ \bibinfo {pages} {054423}
  (\bibinfo {year} {2018})}\BibitemShut {NoStop}%
\bibitem [{\citenamefont {Gruber}\ \emph {et~al.}(2002)\citenamefont {Gruber},
  \citenamefont {Justice}, \citenamefont {Edgar F.~Westrum},\ and\
  \citenamefont {Zandi}}]{Gru-2002}%
  \BibitemOpen
  \bibfield  {author} {\bibinfo {author} {\bibfnamefont {J.~B.}\ \bibnamefont
  {Gruber}}, \bibinfo {author} {\bibfnamefont {B.~H.}\ \bibnamefont {Justice}},
  \bibinfo {author} {\bibfnamefont {J.}~\bibnamefont {Edgar F.~Westrum}}, \
  and\ \bibinfo {author} {\bibfnamefont {B.}~\bibnamefont {Zandi}},\ }\href
  {\doibase https://doi.org/10.1006/jcht.2001.0860} {\bibfield  {journal}
  {\bibinfo  {journal} {The Journal of Chemical Thermodynamics}\ }\textbf
  {\bibinfo {volume} {34}},\ \bibinfo {pages} {457 } (\bibinfo {year}
  {2002})}\BibitemShut {NoStop}%
\bibitem [{\citenamefont {Thalmeier}\ and\ \citenamefont
  {Fulde}(1982)}]{Tha-1982}%
  \BibitemOpen
  \bibfield  {author} {\bibinfo {author} {\bibfnamefont {P.}~\bibnamefont
  {Thalmeier}}\ and\ \bibinfo {author} {\bibfnamefont {P.}~\bibnamefont
  {Fulde}},\ }\href {\doibase 10.1103/PhysRevLett.49.1588} {\bibfield
  {journal} {\bibinfo  {journal} {Phys. Rev. Lett.}\ }\textbf {\bibinfo
  {volume} {49}},\ \bibinfo {pages} {1588} (\bibinfo {year}
  {1982})}\BibitemShut {NoStop}%
\bibitem [{\citenamefont {Loewenhaupt}\ and\ \citenamefont
  {Witte}(2003)}]{Loe-2003}%
  \BibitemOpen
  \bibfield  {author} {\bibinfo {author} {\bibfnamefont {M.}~\bibnamefont
  {Loewenhaupt}}\ and\ \bibinfo {author} {\bibfnamefont {U.}~\bibnamefont
  {Witte}},\ }\href {http://stacks.iop.org/0953-8984/15/i=5/a=307} {\bibfield
  {journal} {\bibinfo  {journal} {Journal of Physics: Condensed Matter}\
  }\textbf {\bibinfo {volume} {15}},\ \bibinfo {pages} {S519} (\bibinfo {year}
  {2003})}\BibitemShut {NoStop}%
\bibitem [{\citenamefont {Gaudet}\ \emph {et~al.}(2018)\citenamefont {Gaudet},
  \citenamefont {Hallas}, \citenamefont {Buhariwalla}, \citenamefont {Sala},
  \citenamefont {Stone}, \citenamefont {Tachibana}, \citenamefont {Baroudi},
  \citenamefont {Cava},\ and\ \citenamefont {Gaulin}}]{Gau-2018}%
  \BibitemOpen
  \bibfield  {author} {\bibinfo {author} {\bibfnamefont {J.}~\bibnamefont
  {Gaudet}}, \bibinfo {author} {\bibfnamefont {A.~M.}\ \bibnamefont {Hallas}},
  \bibinfo {author} {\bibfnamefont {C.~R.~C.}\ \bibnamefont {Buhariwalla}},
  \bibinfo {author} {\bibfnamefont {G.}~\bibnamefont {Sala}}, \bibinfo {author}
  {\bibfnamefont {M.~B.}\ \bibnamefont {Stone}}, \bibinfo {author}
  {\bibfnamefont {M.}~\bibnamefont {Tachibana}}, \bibinfo {author}
  {\bibfnamefont {K.}~\bibnamefont {Baroudi}}, \bibinfo {author} {\bibfnamefont
  {R.~J.}\ \bibnamefont {Cava}}, \ and\ \bibinfo {author} {\bibfnamefont
  {B.~D.}\ \bibnamefont {Gaulin}},\ }\href {\doibase
  10.1103/PhysRevB.98.014419} {\bibfield  {journal} {\bibinfo  {journal} {Phys.
  Rev. B}\ }\textbf {\bibinfo {volume} {98}},\ \bibinfo {pages} {014419}
  (\bibinfo {year} {2018})}\BibitemShut {NoStop}%
\bibitem [{\citenamefont {Adroja}\ \emph {et~al.}(2012)\citenamefont {Adroja},
  \citenamefont {del Moral}, \citenamefont {de~la Fuente}, \citenamefont
  {Fraile}, \citenamefont {Goremychkin}, \citenamefont {Taylor}, \citenamefont
  {Hillier},\ and\ \citenamefont {Fernandez-Alonso}}]{Adj-2012}%
  \BibitemOpen
  \bibfield  {author} {\bibinfo {author} {\bibfnamefont {D.~T.}\ \bibnamefont
  {Adroja}}, \bibinfo {author} {\bibfnamefont {A.}~\bibnamefont {del Moral}},
  \bibinfo {author} {\bibfnamefont {C.}~\bibnamefont {de~la Fuente}}, \bibinfo
  {author} {\bibfnamefont {A.}~\bibnamefont {Fraile}}, \bibinfo {author}
  {\bibfnamefont {E.~A.}\ \bibnamefont {Goremychkin}}, \bibinfo {author}
  {\bibfnamefont {J.~W.}\ \bibnamefont {Taylor}}, \bibinfo {author}
  {\bibfnamefont {A.~D.}\ \bibnamefont {Hillier}}, \ and\ \bibinfo {author}
  {\bibfnamefont {F.}~\bibnamefont {Fernandez-Alonso}},\ }\href {\doibase
  10.1103/PhysRevLett.108.216402} {\bibfield  {journal} {\bibinfo  {journal}
  {Phys. Rev. Lett.}\ }\textbf {\bibinfo {volume} {108}},\ \bibinfo {pages}
  {216402} (\bibinfo {year} {2012})}\BibitemShut {NoStop}%
\bibitem [{\citenamefont {Heyen}\ \emph {et~al.}(1991)\citenamefont {Heyen},
  \citenamefont {Wegerer}, \citenamefont {Sch\"onherr},\ and\ \citenamefont
  {Cardona}}]{Hey-1991}%
  \BibitemOpen
  \bibfield  {author} {\bibinfo {author} {\bibfnamefont {E.~T.}\ \bibnamefont
  {Heyen}}, \bibinfo {author} {\bibfnamefont {R.}~\bibnamefont {Wegerer}},
  \bibinfo {author} {\bibfnamefont {E.}~\bibnamefont {Sch\"onherr}}, \ and\
  \bibinfo {author} {\bibfnamefont {M.}~\bibnamefont {Cardona}},\ }\href
  {\doibase 10.1103/PhysRevB.44.10195} {\bibfield  {journal} {\bibinfo
  {journal} {Phys. Rev. B}\ }\textbf {\bibinfo {volume} {44}},\ \bibinfo
  {pages} {10195} (\bibinfo {year} {1991})}\BibitemShut {NoStop}%
\bibitem [{\citenamefont {Avisar}\ and\ \citenamefont
  {Livneh}(2016)}]{Avi-2016}%
  \BibitemOpen
  \bibfield  {author} {\bibinfo {author} {\bibfnamefont {D.}~\bibnamefont
  {Avisar}}\ and\ \bibinfo {author} {\bibfnamefont {T.}~\bibnamefont
  {Livneh}},\ }\href {\doibase https://doi.org/10.1016/j.vibspec.2016.05.006}
  {\bibfield  {journal} {\bibinfo  {journal} {Vibrational Spectroscopy}\
  }\textbf {\bibinfo {volume} {86}},\ \bibinfo {pages} {14 } (\bibinfo {year}
  {2016})}\BibitemShut {NoStop}%
\bibitem [{\citenamefont {Gouteron}\ \emph {et~al.}(1981)\citenamefont
  {Gouteron}, \citenamefont {Michel}, \citenamefont {Lejus},\ and\
  \citenamefont {Zarembowitch}}]{Gou-1981}%
  \BibitemOpen
  \bibfield  {author} {\bibinfo {author} {\bibfnamefont {J.}~\bibnamefont
  {Gouteron}}, \bibinfo {author} {\bibfnamefont {D.}~\bibnamefont {Michel}},
  \bibinfo {author} {\bibfnamefont {A.}~\bibnamefont {Lejus}}, \ and\ \bibinfo
  {author} {\bibfnamefont {J.}~\bibnamefont {Zarembowitch}},\ }\href {\doibase
  https://doi.org/10.1016/0022-4596(81)90058-X} {\bibfield  {journal} {\bibinfo
   {journal} {Journal of Solid State Chemistry}\ }\textbf {\bibinfo {volume}
  {38}},\ \bibinfo {pages} {288 } (\bibinfo {year} {1981})}\BibitemShut
  {NoStop}%
\bibitem [{\citenamefont {Zarembowitch}\ \emph {et~al.}()\citenamefont
  {Zarembowitch}, \citenamefont {Gouteron},\ and\ \citenamefont
  {Lejus}}]{Zar-1979}%
  \BibitemOpen
  \bibfield  {author} {\bibinfo {author} {\bibfnamefont {J.}~\bibnamefont
  {Zarembowitch}}, \bibinfo {author} {\bibfnamefont {J.}~\bibnamefont
  {Gouteron}}, \ and\ \bibinfo {author} {\bibfnamefont {A.~M.}\ \bibnamefont
  {Lejus}},\ }\href {\doibase 10.1002/pssb.2220940128} {\bibfield  {journal}
  {\bibinfo  {journal} {physica status solidi (b)}\ }\textbf {\bibinfo {volume}
  {94}},\ \bibinfo {pages} {249}},\ \Eprint
  {http://arxiv.org/abs/https://onlinelibrary.wiley.com/doi/pdf/10.1002/pssb.2220940128}
  {https://onlinelibrary.wiley.com/doi/pdf/10.1002/pssb.2220940128}
  \BibitemShut {NoStop}%
\bibitem [{\citenamefont {Pierre}\ \emph {et~al.}(1984)\citenamefont {Pierre},
  \citenamefont {Galera},\ and\ \citenamefont {Bouillot}}]{Pie-1984}%
  \BibitemOpen
  \bibfield  {author} {\bibinfo {author} {\bibfnamefont {J.}~\bibnamefont
  {Pierre}}, \bibinfo {author} {\bibfnamefont {R.}~\bibnamefont {Galera}}, \
  and\ \bibinfo {author} {\bibfnamefont {J.}~\bibnamefont {Bouillot}},\ }\href
  {\doibase https://doi.org/10.1016/0304-8853(84)90301-9} {\bibfield  {journal}
  {\bibinfo  {journal} {Journal of Magnetism and Magnetic Materials}\ }\textbf
  {\bibinfo {volume} {42}},\ \bibinfo {pages} {139 } (\bibinfo {year}
  {1984})}\BibitemShut {NoStop}%
\bibitem [{\citenamefont {Osborn}\ \emph {et~al.}(1987)\citenamefont {Osborn},
  \citenamefont {Loewenhaupt}, \citenamefont {Rainford},\ and\ \citenamefont
  {Stirling}}]{Osb-1987}%
  \BibitemOpen
  \bibfield  {author} {\bibinfo {author} {\bibfnamefont {R.}~\bibnamefont
  {Osborn}}, \bibinfo {author} {\bibfnamefont {M.}~\bibnamefont {Loewenhaupt}},
  \bibinfo {author} {\bibfnamefont {B.}~\bibnamefont {Rainford}}, \ and\
  \bibinfo {author} {\bibfnamefont {W.}~\bibnamefont {Stirling}},\ }in\ \href
  {\doibase https://doi.org/10.1016/B978-1-4832-2948-5.50024-1} {\emph
  {\bibinfo {booktitle} {Anomalous Rare Earths and Actinides}}},\ \bibinfo
  {editor} {edited by\ \bibinfo {editor} {\bibfnamefont {J.}~\bibnamefont
  {Boucherle}}, \bibinfo {editor} {\bibfnamefont {J.}~\bibnamefont {Flouquet}},
  \bibinfo {editor} {\bibfnamefont {C.}~\bibnamefont {Lacroix}}, \ and\
  \bibinfo {editor} {\bibfnamefont {J.}~\bibnamefont {Rossat-Mignod}}}\
  (\bibinfo  {publisher} {Elsevier},\ \bibinfo {year} {1987})\ pp.\ \bibinfo
  {pages} {70 -- 72}\BibitemShut {NoStop}%
\bibitem [{\citenamefont {Schedler}\ \emph {et~al.}(2003)\citenamefont
  {Schedler}, \citenamefont {Rotter}, \citenamefont {Witte}, \citenamefont
  {Loewenhaupt},\ and\ \citenamefont {Schmidt}}]{Sch-2003}%
  \BibitemOpen
  \bibfield  {author} {\bibinfo {author} {\bibfnamefont {R.}~\bibnamefont
  {Schedler}}, \bibinfo {author} {\bibfnamefont {M.}~\bibnamefont {Rotter}},
  \bibinfo {author} {\bibfnamefont {U.}~\bibnamefont {Witte}}, \bibinfo
  {author} {\bibfnamefont {M.}~\bibnamefont {Loewenhaupt}}, \ and\ \bibinfo
  {author} {\bibfnamefont {W.}~\bibnamefont {Schmidt}},\ }\href@noop {}
  {\bibfield  {journal} {\bibinfo  {journal} {Acta Physica Polonica B - ACTA
  PHYS POL B}\ }\textbf {\bibinfo {volume} {34}} (\bibinfo {year}
  {2003})}\BibitemShut {NoStop}%
\bibitem [{\citenamefont {Hillier}\ \emph {et~al.}(2012)\citenamefont
  {Hillier}, \citenamefont {Adroja}, \citenamefont {Manuel}, \citenamefont
  {Anand}, \citenamefont {Taylor}, \citenamefont {McEwen}, \citenamefont
  {Rainford},\ and\ \citenamefont {Koza}}]{Hil-2012}%
  \BibitemOpen
  \bibfield  {author} {\bibinfo {author} {\bibfnamefont {A.~D.}\ \bibnamefont
  {Hillier}}, \bibinfo {author} {\bibfnamefont {D.~T.}\ \bibnamefont {Adroja}},
  \bibinfo {author} {\bibfnamefont {P.}~\bibnamefont {Manuel}}, \bibinfo
  {author} {\bibfnamefont {V.~K.}\ \bibnamefont {Anand}}, \bibinfo {author}
  {\bibfnamefont {J.~W.}\ \bibnamefont {Taylor}}, \bibinfo {author}
  {\bibfnamefont {K.~A.}\ \bibnamefont {McEwen}}, \bibinfo {author}
  {\bibfnamefont {B.~D.}\ \bibnamefont {Rainford}}, \ and\ \bibinfo {author}
  {\bibfnamefont {M.~M.}\ \bibnamefont {Koza}},\ }\href {\doibase
  10.1103/PhysRevB.85.134405} {\bibfield  {journal} {\bibinfo  {journal} {Phys.
  Rev. B}\ }\textbf {\bibinfo {volume} {85}},\ \bibinfo {pages} {134405}
  (\bibinfo {year} {2012})}\BibitemShut {NoStop}%
\bibitem [{\citenamefont {Smidman}\ \emph {et~al.}(2013)\citenamefont
  {Smidman}, \citenamefont {Adroja}, \citenamefont {Hillier}, \citenamefont
  {Chapon}, \citenamefont {Taylor}, \citenamefont {Anand}, \citenamefont
  {Singh}, \citenamefont {Lees}, \citenamefont {Goremychkin}, \citenamefont
  {Koza}, \citenamefont {Krishnamurthy}, \citenamefont {Paul},\ and\
  \citenamefont {Balakrishnan}}]{Smi-2013}%
  \BibitemOpen
  \bibfield  {author} {\bibinfo {author} {\bibfnamefont {M.}~\bibnamefont
  {Smidman}}, \bibinfo {author} {\bibfnamefont {D.~T.}\ \bibnamefont {Adroja}},
  \bibinfo {author} {\bibfnamefont {A.~D.}\ \bibnamefont {Hillier}}, \bibinfo
  {author} {\bibfnamefont {L.~C.}\ \bibnamefont {Chapon}}, \bibinfo {author}
  {\bibfnamefont {J.~W.}\ \bibnamefont {Taylor}}, \bibinfo {author}
  {\bibfnamefont {V.~K.}\ \bibnamefont {Anand}}, \bibinfo {author}
  {\bibfnamefont {R.~P.}\ \bibnamefont {Singh}}, \bibinfo {author}
  {\bibfnamefont {M.~R.}\ \bibnamefont {Lees}}, \bibinfo {author}
  {\bibfnamefont {E.~A.}\ \bibnamefont {Goremychkin}}, \bibinfo {author}
  {\bibfnamefont {M.~M.}\ \bibnamefont {Koza}}, \bibinfo {author}
  {\bibfnamefont {V.~V.}\ \bibnamefont {Krishnamurthy}}, \bibinfo {author}
  {\bibfnamefont {D.~M.}\ \bibnamefont {Paul}}, \ and\ \bibinfo {author}
  {\bibfnamefont {G.}~\bibnamefont {Balakrishnan}},\ }\href {\doibase
  10.1103/PhysRevB.88.134416} {\bibfield  {journal} {\bibinfo  {journal} {Phys.
  Rev. B}\ }\textbf {\bibinfo {volume} {88}},\ \bibinfo {pages} {134416}
  (\bibinfo {year} {2013})}\BibitemShut {NoStop}%
\bibitem [{\citenamefont {Hayes}\ and\ \citenamefont {Loudon}(1978)}]{Hay-Lou}%
  \BibitemOpen
  \bibfield  {author} {\bibinfo {author} {\bibfnamefont {W.}~\bibnamefont
  {Hayes}}\ and\ \bibinfo {author} {\bibfnamefont {R.}~\bibnamefont {Loudon}},\
  }\href@noop {} {\emph {\bibinfo {title} {Scattering of Light by Crystals}}}\
  (\bibinfo  {publisher} {Dover Publications},\ \bibinfo {year}
  {1978})\BibitemShut {NoStop}%
\bibitem [{\citenamefont {Lushnikov}\ \emph {et~al.}(2004)\citenamefont
  {Lushnikov}, \citenamefont {Gvasaliya},\ and\ \citenamefont
  {Katiyar}}]{Lus-2004}%
  \BibitemOpen
  \bibfield  {author} {\bibinfo {author} {\bibfnamefont {S.}~\bibnamefont
  {Lushnikov}}, \bibinfo {author} {\bibfnamefont {S.}~\bibnamefont
  {Gvasaliya}}, \ and\ \bibinfo {author} {\bibfnamefont {R.~S.}\ \bibnamefont
  {Katiyar}},\ }\href {\doibase 10.1103/PhysRevB.70.172101} {\bibfield
  {journal} {\bibinfo  {journal} {Phys. Rev. B}\ }\textbf {\bibinfo {volume}
  {70}},\ \bibinfo {pages} {172101} (\bibinfo {year} {2004})}\BibitemShut
  {NoStop}%
\bibitem [{\citenamefont {Kumar}\ \emph {et~al.}(2011)\citenamefont {Kumar},
  \citenamefont {Scott},\ and\ \citenamefont {Katiyar}}]{Ash-2011}%
  \BibitemOpen
  \bibfield  {author} {\bibinfo {author} {\bibfnamefont {A.}~\bibnamefont
  {Kumar}}, \bibinfo {author} {\bibfnamefont {J.~F.}\ \bibnamefont {Scott}}, \
  and\ \bibinfo {author} {\bibfnamefont {R.~S.}\ \bibnamefont {Katiyar}},\
  }\href {\doibase 10.1063/1.3624845} {\bibfield  {journal} {\bibinfo
  {journal} {Applied Physics Letters}\ }\textbf {\bibinfo {volume} {99}},\
  \bibinfo {pages} {062504} (\bibinfo {year} {2011})},\ \Eprint
  {http://arxiv.org/abs/https://doi.org/10.1063/1.3624845}
  {https://doi.org/10.1063/1.3624845} \BibitemShut {NoStop}%
\bibitem [{\citenamefont {Wemple}\ and\ \citenamefont
  {DiDomenico}(1969)}]{Wem-1969}%
  \BibitemOpen
  \bibfield  {author} {\bibinfo {author} {\bibfnamefont {S.~H.}\ \bibnamefont
  {Wemple}}\ and\ \bibinfo {author} {\bibfnamefont {M.}~\bibnamefont
  {DiDomenico}},\ }in\ \href@noop {} {\emph {\bibinfo {booktitle} {Light
  Scattering Spectra of Solids}}},\ \bibinfo {editor} {edited by\ \bibinfo
  {editor} {\bibfnamefont {G.~B.}\ \bibnamefont {Wright}}}\ (\bibinfo
  {publisher} {Springer Berlin Heidelberg},\ \bibinfo {address} {Berlin,
  Heidelberg},\ \bibinfo {year} {1969})\ pp.\ \bibinfo {pages}
  {65--74}\BibitemShut {NoStop}%
\end{thebibliography}

%

\end{document}